\documentclass{nature}

\usepackage{blindtext}
\usepackage[T1]{fontenc}
\usepackage{CJKutf8}
\usepackage{centernot}
\usepackage{graphicx}
\usepackage{amsmath,bbold}
\usepackage{times}
\usepackage{amssymb}
\usepackage{mathrsfs}
\usepackage{chemarr}
\usepackage{color}
\usepackage{url}
\usepackage{version}
\usepackage[hidelinks]{hyperref}
\usepackage{mwe,tikz}
\usepackage[percent]{overpic}
\usepackage{bm}
\usepackage[export]{adjustbox}
\definecolor{linkcolor}{rgb}{0,0,0.6}

\newcommand{\br}{{\bf r}}

\newcommand{\rr}{\bm{\mathrm{r}}}

\newcommand{\uu}{\bm{\mathrm{u}}}
\newcommand{\mm}{\bm{\mathrm{m}}}
\newcommand{\vv}{\bm{\mathrm{v}}}
\newcommand{\Q}{\bm{\mathrm{Q}}}
\newcommand{\J}{\bm{\mathrm{J}}}

\newcommand{\gLG}{\gamma_{\rm LG}}
\newcommand{\gGS}{\gamma_{\rm GS}}
\newcommand{\gLS}{\gamma_{\rm LS}}

\newcommand\diff[2]{\frac{d #1}{d #2}}

\newcommand\pdiff[2]{\frac{\partial #1}{\partial #2}}

\newcommand\red[1]{{\color{red}#1}}

\usepackage{lipsum}
\usetikzlibrary{patterns}
\usetikzlibrary{shapes}
\graphicspath{{./Figs/}}
\newcommand\rp{{\bf r}_{\rm p}}

\usepackage{caption}
\usepackage{tocloft}
\newcounter{savedbib}

\title{Wetting by active fluids}

\author{Yongfeng Zhao ({\begin{CJK*}{UTF8}{gbsn}赵永峰\end{CJK*}})$^{1,2,\dagger}$, Ruben Zakine$^{3}$, Adrian Daerr$^{4}$, Yariv Kafri$^5$, Julien Tailleur$^{6}$, Frédéric van Wijland$^{4}$}

\begin{document}
\addtocontents{toc}{\protect\setcounter{tocdepth}{-1}}
\maketitle 

\begin{affiliations}
\item{Center for Soft Condensed Matter Physics and Interdisciplinary Research \& School of Physical Science and Technology, Soochow University, 215006 Suzhou, China}
\item{School of Physics and Astronomy and Institute of Natural Sciences, Shanghai Jiao Tong University, Shanghai 200240, China}
\item{Laboratoire d'Hydrodynamique, CNRS, École polytechnique, Institut Polytechnique de Paris, 91120 Palaiseau, France}
\item{Universit\'e Paris Cité, Laboratoire Mati\`ere et Syst\`emes Complexes (MSC), UMR 7057 CNRS, F-75205 Paris, France}
\item{Department of Physics, Technion -- Israel Institute of Technology, Haifa 3200003, Israel}
\item Department of Physics, Massachusetts Institute of Technology, Cambridge, Massachusetts 02139, USA
\end{affiliations}

\noindent$^\dagger$ Corresponding author: {\small
  \texttt{yfzhao2021@suda.edu.cn}%
}

\date{\today}

\begin{abstract}
The Young-Dupré equation is a cornerstone of the theory of capillary and wetting phenomena that governs the shape of droplets adsorbed on surfaces in equilibrium. But many living and synthetic materials, such as swarming bacteria and active colloids, are composed of self-propelled particles that are inherently out of thermal equilibrium. The description of the wetting of surfaces by such active fluids thus requires a new framework. Here, we develop an analogue to the Young-Dupré equation for systems made of self-propelled particles. A key step is to define the liquid-gas surface tension of active fluids as the force exerted along the interface, which we show from first principles to be negative, even when active materials separate into stable liquid and gas phases. The active Young-Dupré equation explains why partial wetting appears in simulations where the surface tensions do not balance and reveals the underlying feedback mechanism: the interface is stable only because of steady flows, which are themselves generated by the parity-symmetry-breaking interface. Unlike in passive fluids, where the droplets are scale-free, this feedback loop selects the sizes and shapes of adsorbed droplets in active materials. Our results outline a framework for understanding how active matter wets surfaces.
\end{abstract}

\begin{figure}
  \centering
  \includegraphics{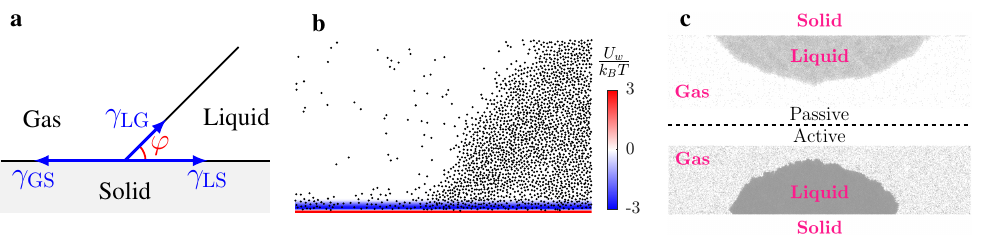}
    
  \caption{{\bf Wetting of a solid surface by a
macroscopic liquid droplet.} {\bf a}, In equilibrium, when a macroscopic droplet wets a
    solid surface, the contact angle $\varphi$ satisfies the
    Young-Dupré equation~\eqref{eq:YD}. The arrows define the
    directions of the macroscopic forces exerted on the contact line
    for positive surface tensions. {\bf b}, Snapshot of a
    phase-separating passive system partially wetting  a solid
    confining wall, showing that, at the particle scale, the liquid-gas interface is smeared over a finite microscopic  length. {\bf c}, Snapshots of passive (top) and active
      (bottom) droplets wetting solid surfaces, showing the apparent
      similarities between the two cases.}
    \label{newfig1}
\end{figure}

The Young-Dupr\'{e} equation describes the equilibrium wetting of a
macroscopic liquid droplet on a solid surface as illustrated
in Fig.~\ref{newfig1}a. It determines the contact angle $\varphi$ between the
liquid-gas interface and the solid surface through
\begin{equation} 
\gamma_{\rm GS}-\gamma_{\rm LS}=\gamma_{\rm LG}\cos\varphi\;,\label{eq:YD}
\end{equation}
where $\gamma_{XY}$ is the surface tension between phases $X$ and $Y$.
This equation is a cornerstone of the theory of capillary
phenomena~\cite{de1985wetting}, with implications ranging from wetting
transitions~\cite{bonn2009} to finite relaxation times in liquid
entrainment~\cite{snoeijer2006avoided}. 
A remarkable property of the Young-Dupré equation is that it endows interfaces with mechanical properties:
Eq.~\eqref{eq:YD} can be read as a force balance condition along the direction tangential to the solid substrate as shown in Fig.~\ref{newfig1}a. 
This macroscopic description emerges despite the fact that interfaces are smeared fluctuating objects at the particle scale where the forces are
actually exerted, see Fig.~\ref{newfig1}b. 
In equilibrium, an important challenge has thus
been to bridge these macroscopic `forces' to their microscopic origin,
which results from the interplay between particle interactions and
thermal fluctuations~\cite{tarazona1981}. Meeting this challenge has
led to a comprehensive understanding of interface physics in
equilibrium~\cite{de1985wetting}.

Reaching a comparable state of understanding in active matter is an
important open challenge due to the flurry of interfacial phenomena
observed in active systems---from the growth of cellular
tissues~\cite{sepulveda2013collective} and bacterial
colonies~\cite{hallatschek2007genetic} to the cohesion of flocks of
birds~\cite{bialek2012statistical} or colloidal
rollers~\cite{bricard2013emergence}. The quest for a micro-to-macro
theory of interface dynamics in active systems has thus attracted an
upsurge of attention over the past few
years~\cite{joanny2012drop,bialke15,marconi2016pressure,paliwal17,patch2018,wittmann2019pressure,omar2020,wysocki2020,fausti2021,neta2021,adkins2022dynamics,caballero2022activity,tayar2023controlling,rojasvega2023,li2023surface,mangeat2024pre,turci2024partial}. So
far, however, a comprehensive theory of interface mechanics and
equilibria is lacking for active systems, which raises the question as
to whether a counterpart to the Young-Dupr\'e equation exists in this
context. In this work, we fill this gap for self-propelled particles interacting via pairwise forces.

\section{Active surface (ex)tension: when interfaces push on confining walls.}
An inspiring and fruitful starting point consists in directly
exporting equilibrium results to active materials. However, an
important difficulty arises from the continuous injection of energy
and momentum in the steady states of active systems, which has made the very definition of surface tension a challenging task. While a mechanical definition of fluid-solid surface tension has been proposed recently~\cite{zakine2020}, the liquid-gas surface tension has
been the topic of much
controversy~\cite{bialke15,hermann2019,omar2020,lauersdorf2021phase,li2023surface,turci2024partial}.
Progress has been made on this question by considering the workhorse
models of self-propelled particles interacting via pairwise
forces~\cite{fily2012athermal}. These systems undergo a Motility-Induced Phase Separation (MIPS) between
a dilute and a dense phase in the presence of repulsive forces at
large enough density and propulsion
speed~\cite{fily2012athermal,cates2015motility}. 
Similarly to the passive case, partial wetting is then observed in the presence of
a confining solid surface, as
illustrated in Fig.~\ref{newfig1}c. 
(All our simulations are detailed in Methods.) 

For
such models, an interesting inroad into surface tension was recently
proposed~\cite{bialke15} by assuming that the equilibrium expression
of a liquid-gas surface
tension~\cite{kirkwood49,navascues77} also holds for
active interfaces:
\begin{equation}
\gamma_{\rm LG}=\int_{\br_{\rm G}}^{\br_{\rm L}}dr_\perp \, [p_\perp-p_\parallel(r_\perp)]\;, \label{eq:gamma_LG_speck}
\end{equation}
where $\br_{\rm G}$ and $\br_{\rm L}$ are positions in the bulk gas
and liquid phases, respectively, and $r_\perp$ is a coordinate normal
to the interface. In Eq.~\eqref{eq:gamma_LG_speck}, $p_\perp$ is the
pressure normal to the interface while $p_\parallel$ is the
tangential pressure. However,
Eq.~\eqref{eq:gamma_LG_speck} has been challenged for a number of
reasons~\cite{hermann2019,
  omar2020,lauersdorf2021phase,li2023surface}, starting from the resulting
negative value of $\gamma_{\rm LG}$ that, in passive systems, signals an unstable
interface. Furthermore, a \textit{direct} measurement of a force
resulting from the surface tension defined in
Eq.~\eqref{eq:gamma_LG_speck} is lacking. This has left room for
alternative definitions of $\gamma_{\rm LG}$ that result in a positive
value~\cite{hermann2019,omar2020,li2023surface,turci2024partial}. We note that active field theories, which have proven useful to account for the self-organization of active-matter systems~\cite{marchetti2013hydrodynamics,cates2024active}, are here of little help: since active fluids do not admit a free energy, there is no ambiguity-free way to couple external potentials and density fields directly at the coarse-grained scale. This prohibits a top-down approach to active mechanics, which would start from an active field theory and predict the force exerted by the corresponding active system on its environment.

Instead, starting from first principles, we show that Eq.~\eqref{eq:gamma_LG_speck}
can be derived from microscopics, leading to a surface tension $\gamma_{\rm LG}$ that is a measurable force exerted along interfaces, and that this force is indeed negative. To do so, we consider self-propelled particles
interacting via pairwise forces, such that the dynamics of particle
$i$ at position $\rr_i$ and orientation $\uu_i$ reads:
\begin{equation}
\dot{\rr}_i=v_0\uu_i-\mu\sum_{j\neq i}\nabla_{\rr_i}U(\rr_i-\rr_j)-\mu\nabla_{\rr_i}U_w(\rr_i)\;, \label{eq:eomr}
\end{equation}
where $v_0$ is the particle propulsion speed, $\mu$ its
mobility and $U$ a repulsive pair potential. When the system is in presence of a confining
solid boundary, we model the latter as a repulsive potential
$U_w(\rr)$, following the seminal work of Navascu\'es and
Berry~\cite{navascues77}. For the sake of concreteness, we consider
overdamped active Brownian particles (ABPs) in two
dimensions~\cite{fily2012athermal} so that
$\uu_i=(\cos\theta_i,\sin\theta_i)$ and the orientations evolve as
$\dot{\theta}_i=\sqrt{2/\tau_p}\,\xi_i$, with $\tau_p$ the persistence
time and $\xi_i$ a unit-variance centered Gaussian white noise. We
note, however, that our results should straightforwardly extend to $d=3$ dimensions and to
the broader class of systems whose particle-propulsion forces have an
autonomous dynamics, which warrants the existence of an equation of
state for the pressure~\cite{fily17,das2019local}. As an example, we generalize our results to active
Ornstein-Uhlenbeck particles~\cite{sepulveda2013collective,Szamel2014PRE} in SM Section~\ref{SM:sec:AOUP}.

For the class of models we consider, the sole hydrodynamic field is the conserved density
$\rho(\rr,t)=\langle\sum_i\delta(\rr-\rr_i(t))\rangle$, whose dynamics reads~\cite{solon2015natphys,solon2018generalized}: 
\begin{equation}
\dot \rho = -\nabla \cdot \J;\quad\text{with}\quad
\mu^{-1} \J= \nabla\cdot\bm{\sigma}-\rho\nabla_{\rr}U_w(\rr)\;,\label{eq:ss_flow}
\end{equation}
where $\J(\rr)=\langle\sum_i\dot{\rr}_i\delta(\rr-\rr_i(t))\rangle$ is
the particle current (See SM Section~\ref{SM:sec:stress}). In Eq.~\eqref{eq:ss_flow},
$\bm{\sigma}= \bm \sigma^{\rm IK}+\bm \sigma^{\rm a}$ acquires the
physical meaning of a stress tensor in the steady state, so that the pressure in a bulk phase is $p=-{\rm Tr} \bm{\sigma}/2$~\cite{solon2015pressure}. The stress tensor comprises
two
contributions~\cite{yang2014,takatori2014,solon2015pressure,solon2018generalized}:
$\bm \sigma^{\rm IK}$ is the standard Irving-Kirkwood contribution
stemming from the pairwise interactions between the particles and $\bm
\sigma^{\rm a}$ is a contribution due to the active
forces. 
Equation~\eqref{eq:ss_flow} is well established~\cite{solon2015pressure,solon2018generalized,zakine2020}: it relates the drag density exerted on the
particles, ${\bf f}_d\equiv -\mu^{-1}\J$, and the stress tensor
$\bm\sigma$ to any external force applied to the system; it is the starting point of our theory.

\begin{figure*}
\begin{center}
  \includegraphics{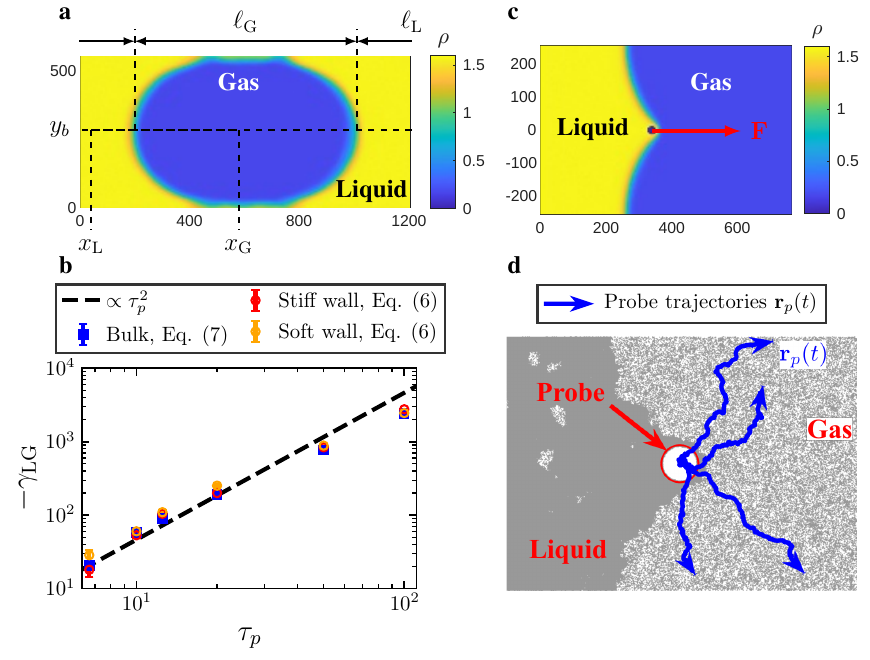}
  \end{center}
\caption{{\bf Surface (ex)tension: when active fluids push along interfaces.} \textbf{a}, Phase-separated system comprising Active
  Brownian Particles, with periodic boundary conditions along $\hat x$
  and confining walls at the bottom and top of the system. 
  \textbf{b}, The measurements of $\gamma_{\rm LG}$ from the wall force through Eq.~\eqref{eq:Ftot2}, $F_{w,\rm tot} = -n_{{\rm LG}} \gamma_{\rm LG}+ \ell_{\rm G} p_{\rm G} + \ell_{\rm L} p_{\rm L}$, and its bulk counterpart Eq.~\eqref{eq:gamma_LG}, $\gamma_{\rm LG}=\int_{x_{\rm L}}^{x_{\rm G}}dx\,[\sigma_{yy}(x,y_b)-\sigma_{xx}(x,y_b)]$, agree
  quantitatively over a large range of persistence times. 
  The scaling $\gamma_{\rm LG}
  \propto \tau_p^2$ inherited from the active stress tensor, $\bm
  \sigma^{\rm a}\propto v_0^2\tau_p$, and the width of the
  interface region, $v_0\tau_p$, is confirmed by our numerics. Error bars represent three times the standard error of the mean (s.e.m.), estimated from $M\geq 2138$ independent block samples (see Methods). \textbf{c}, Wilhelmy-plate setup: Average density profile once the plate has been pulled out of the liquid. The average force $\bf F$ exerted on the immersed plate pushes the probe out of the liquid. \textbf{d}, When the plate is released, it moves due to its interaction
  with the active particles. Four representative plate trajectories all lead to the expulsion of the plate, at odds
  with the equilibrium case in which the plate would be pulled inside
  the liquid phase due to the positive liquid-gas surface
  tension. See also SM Movie~1.}\label{newfig2}
\end{figure*}

To derive the liquid-gas surface tension from first principles, we consider a phase-separated system in a rectangular box, with confining walls along the $\hat y$ direction at the top and bottom of the system and periodic boundary conditions along the horizontal $\hat x$ direction, as shown in Fig.~\ref{newfig2}a.  
To endow the
definition~\eqref{eq:gamma_LG_speck} with a proper mechanical
interpretation, we compute the
total force exerted on the upper wall by the particles, $F_{w, \rm tot}=\int_0^{L_x}dx\int_{y_b}^\infty dy  \rho(x,y) \partial_{y} U_w(y)$ (See SM Section~\ref{SM:sec:LGtension} for further details). At the macroscopic scale,  $F_{w, \rm tot}$ stems both from the contributions of the bulk gas and liquid pressures, $p_{\rm G}$ and $p_{\rm L}$, and from interface contributions that we use to identify the liquid-gas surface tension.
Using Eq.~\eqref{eq:ss_flow}, we integrate $J_y\equiv \J\cdot\hat y$ over the
upper half of the system, for $y>y_b$ and $0\leq x \leq L_x$,
which yields
\begin{equation}\label{eq:Ftot1}
F_{w,\rm tot}=   -\int_{0}^{L_x}\!\!\!\!\! dx' \sigma_{yy}(x',y_b)-\int_{y\geq y_b} \!\!\!\!\!\!\!\! d^2 \br \frac{J_y}{\mu}\;.
\end{equation}
In the steady-state, $\nabla \cdot {\bf J}=0$ implies that the integral of $J_y$ vanishes. 
Furthermore, we can add and subtract $\int_{0}^{L_x}\!  dx'
\sigma_{xx}(x',y_b)=-\ell_{\rm G} p_{\rm G} - \ell_{\rm L} p_{\rm L}$ to Eq.~\eqref{eq:Ftot1}, where $\ell_{{\rm G}/{\rm L}}$ are the total lengths of the gas and liquid
phases, respectively. This leads  to
\begin{equation}\label{eq:Ftot2}
F_{w,\rm tot} =   -n_{{\rm LG}} \gamma_{\rm LG}+ \ell_{\rm G} p_{\rm G} + \ell_{\rm L} p_{\rm L} \;,
\end{equation}
where $n_{\rm LG}$ is the number of liquid-gas
interfaces, allowing us to identify the liquid-gas surface tension
$\gamma_{\rm LG}$ as
\begin{equation}\label{eq:gamma_LG}
\gamma_{\rm LG}=\int_{x_{\rm L}}^{x_{\rm G}}dx\,[\sigma_{yy}(x,y_b)-\sigma_{xx}(x,y_b)]\;.
\end{equation}

Equation~\eqref{eq:Ftot2} is the first central result of our work: it unambiguously defines the liquid-gas surface tension $\gamma_{\rm LG}$ as the contribution to the force exerted on the upper wall due to the presence of interfaces. It matches our equilibrium intuition and Eq.~\eqref{eq:gamma_LG} is indeed equivalent to the one proposed in
Eq.~\eqref{eq:gamma_LG_speck}, thanks to the identification
$p_\perp=-\sigma_{xx}(x,y)$ and $p_\parallel(x,y)=-\sigma_{yy}(x,y)$~\cite{solon2018generalized,speck2020collective, omar2020, langford2023theory}. Since Eq.~\eqref{eq:gamma_LG} provides a way to measure $\gamma_{\rm LG}$ in a phase-separated system without confining walls, the force exerted by the active interface on the upper wall in Eq.~\eqref{eq:Ftot2} is independent of the wall properties: the mechanical surface tension thus obeys an equation of state. 
As shown in Fig.~\ref{newfig2}b,
Eqs.~\eqref{eq:Ftot2} and~\eqref{eq:gamma_LG} agree quantitatively in
our simulations over a wide range of persistence lengths,
independently of the choice of the wall potential. 
The
\textit{negative} value of the surface tension is also confirmed: the
interface exerts a repulsive, outward force on the upper and bottom
walls, in contrast with the equilibrium case. 
Our results thus settle the long-standing debate on the liquid-gas surface tension in active fluids: the mechanical surface tension can be unambiguously defined as Eq.~\eqref{eq:gamma_LG} and measured as Eq.~\eqref{eq:Ftot2}. All other definitions proposed in the literature~\cite{li2023surface,turci2024partial,hermann2019,langford2024mechanics}---which may play an important role in coarsening~\cite{tjhung2018cluster}, interface dynamics~\cite{fausti2021}, or nucleation theory~\cite{cates2023classical}---are not related to mechanics, and should probably not be referred to as  \textit{tensions}.

\section{The Wilhelmy plate experiment}
The negative value of the liquid-gas surface tension
can then be demonstrated using a setup inspired by the Wilhelmy plate
experiment~\cite{volpe2018wilhelmy}. In the latter, a plate---or a
ring in a variant introduced by Pockels and du
Noüy---is inserted into a liquid-gas interface and
slowly pulled out of the liquid. In equilibrium, the positive surface
tension $\gamma_{\rm LG}$ then exerts a restoring force on the plate,
pulling it back into the liquid and resisting the outward motion. The
measurement of the force on the plate can then be used to extract
$\gamma_{\rm LG}$.

Figure~\ref{newfig2}c shows a similar experiment carried out with a disk slowly
pulled out of the liquid phase. As in
the Wilhelmy-plate experiment, a capillary bridge forms, leading to a
deformation of the interface. In our active bath, we then expect
surface tension to induce a net force on the plate, and the negative
value of $\gamma_{\rm LG}$ to translate into an \textit{outward}
force. To test this prediction, we then release our plate at the
configuration shown in Fig.~\ref{newfig2}c, and let its position $\rp(t)$ evolve
according to:
\begin{align}\label{eq:Fwilhelmy}
M\ddot{\bf r}_p=-\sum_i\nabla_{\rp} U_w(\rp-\rr_i)\;,
\end{align} 
where $M$ is the mass of the plate and $U_w(\rr)$ now describes the
interaction potential between the plate and the active particles. Four
representative trajectories are displayed in Fig.~\ref{newfig2}d, which all
show the \textit{expulsion} of the plate out of the liquid phase, in
stark contrast with the passive case. (See also SM Movie~1.) The expulsion of the plate by
the fluid is an unambiguous demonstration that the surface tension of
active fluids, understood in its traditional mechanical sense, is
indeed negative. As shown in SM Section~\ref{SM:sec:wilhelmy}, the outward average force exerted on
the plate can be related quantitatively to the negative liquid-gas
surface tension starting from the mechanical balance
\eqref{eq:ss_flow} between the drag, the divergence of the stress
tensor, and the external force.

\begin{figure*}
\begin{center}
  
  \includegraphics{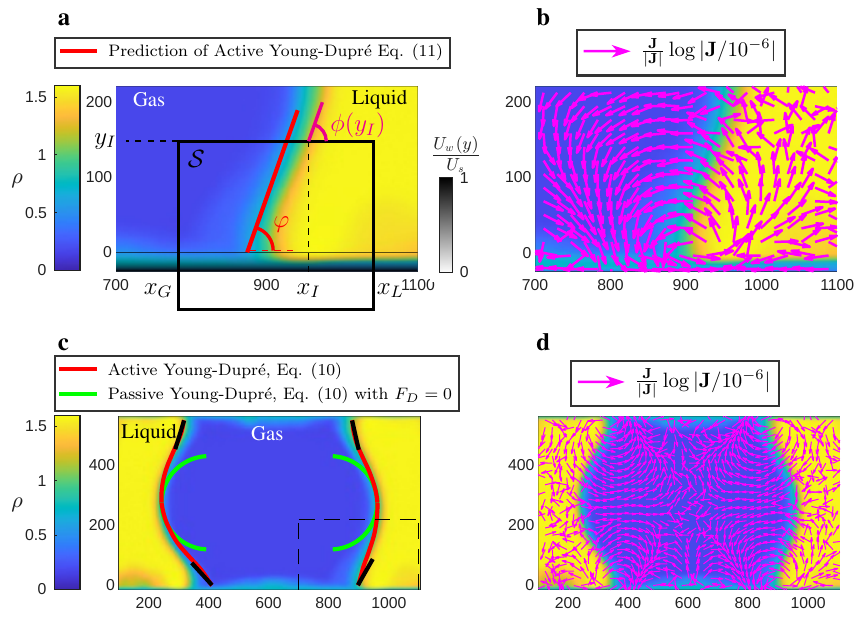}
  \end{center}
  \caption{{\bf The active Young-Dupr\'{e} equation.} \textbf{a}, The density field in the vicinity of the triple point. The red line shows the prediction of the angle given by Eq.~\eqref{eq:yongfengdupre}, $ \gamma_{\rm LG} \cos \varphi=\gamma_{\rm GS}-\gamma_{\rm LS} - F_D(0)$. The repulsive wall potential $U_w$ is encoded using a gray scale, with black corresponding to the value $U_s$ at which the wall stalls the particle propulsion. \textbf{b}, The current field measured in Panel a. The arrow lengths are proportional to $\log|\J/10^{-6}|$. \textbf{c, d} show the full coexistence region with panels a,b corresponding to the black dashed rectangle. The red curves show the interface predicted by Eq.~\eqref{eq:interface_shape}, while the green curve shows the current-free prediction, setting $F_D=0$. Finite wetting is only possible thanks to the stabilizing effect of the vortex flow shown in panel d. The arrow lengths are proportional to $\log|\J/10^{-6}|$. }\label{newfig3}
\end{figure*}

\section{The active Young-Dupr\'{e} equation}

Our derivation of Eq.~\eqref{eq:gamma_LG} as a \textit{bona
  fide} liquid-gas surface tension comes with a challenge: In
the stable partial wetting situation shown in Fig.~\ref{newfig2}a, all surface
tensions can now be measured, but they lead to violations of the
Young-Dupré equation~\eqref{eq:YD}. Indeed, our measurements, detailed in Methods, give $\gamma_{\rm LS}=-631\pm 6$, $\gamma_{\rm GS}=-907\pm5$, and $\gamma_{\rm LG}=-192\pm 3$ in our simulation units
 so:
\begin{equation}\label{eq:YD_imbalance}
  |\gamma_{\rm GS}-\gamma_{\rm LS}|>|\gamma_{\rm LG}|\;.
\end{equation}
According to the passive Young-Dupré Eq.~\eqref{eq:YD}, this mechanical imbalance implies a complete dewetting. 

To account for mechanical equilibrium at the triple
point, we integrate the projection of Eq.~\eqref{eq:ss_flow} along
$\hat x$ over the region $\mathcal{S}$ shown in Fig.~\ref{newfig3}a,
which encompasses the liquid-gas interface up to a height $y_I$. As detailed in SM Section~\ref{SM:sec:ayd}, the generalization of the derivation of Eq.~\eqref{eq:Ftot2} detailed above leads to
\begin{equation}
0=\gamma_{\rm LS}-\gamma_{\rm GS}-y_I\Delta p+\gamma_{\rm LG}\cos\phi(y_I)+F_D(y_I)\;, \label{eq:active_young_dupre}
\end{equation}
where the fluid-solid surface tensions are defined
as~\cite{zakine2020} $\gamma_{\rm GS,LS}=y_I p_{\rm
  G,L}+\int_{-\infty}^{y_I} dy\, \sigma_{xx} (x_{\rm G/L},y)$,
$\phi(y_I)$ is the angle between the liquid-gas interface and the
horizontal axis at the interface coordinate $(x_I,y_I)$, and $F_D=\iint_{\mathcal{S}} \mathbf{f}_d\cdot \hat x\,d^2\br$ is
the total drag exerted on the particles along $\hat x$ in the region
$\mathcal{S}$. We note that, contrary to the case depicted in
Fig.~\ref{newfig1} for which the Young-Dupré law is typically derived, the
liquid-gas interface is curved in our simulations, leading---as in passive systems---to $\Delta p\equiv p_{\rm L}-p_{\rm G}\neq 0$. As we show below, the drag, which is absent in the equilibrium Young-Dupré equation, is the key ingredient that allows for partial wetting by resolving the apparent contradiction found in Eq.~\eqref{eq:YD_imbalance}.

We now use the simulations shown in Fig.~\ref{newfig3} to test the
prediction for the contact angle $\varphi$. To do so, we need to
evaluate each term entering Eq.~\eqref{eq:active_young_dupre}. The
surface tensions can be evaluated in the bulk of the system, where
they are independent of $y_I$. Then, a natural definition of the
contact angle $\varphi$ amounts to taking the limit of $\phi(y_I)$ as $y_I\to
0$. In this limit, the contribution of the pressure difference
vanishes and we recover the active Young-Dupr\'e equation:
\begin{equation}\label{eq:yongfengdupre}
\gamma_{\rm GS}-\gamma_{\rm LS}=\gamma_{\rm LG} \cos \varphi + F_D(0)\;,
\end{equation}
where $F_D(0)$ is the drag experienced by particles in contact with
the wall. Figure~\ref{newfig3}a compares the prediction of the active Young-Dupré
equation with the liquid-gas interface measured in numerical
simulations, leading to a very good agreement between measured and
predicted contact angles. As expected, the macroscopic contact angle
describes the behaviour of the smoothly varying interface at $y \geq
\ell_p$ as it approaches the bottom wall, and it does not resolve the
microscopic structure on the scale $y<\ell_p$. In SM Fig.~\ref{SM:fig:active_YD_sup}, we show that our theory holds for other persistence times and wall
stiffnesses. Figure~\ref{SM:fig:active_YD_interaction_stiffness} also shows how all observables entering the active Young-Dupré equation vary with the particle stiffness, leading to a wetting transition in the limit of soft particles.

Note that our active Young-Dupré equation can be used to predict the full shape of the interface by
writing that
$\cot\phi(y_I)=x_I'(y_I)$. Equation~\eqref{eq:active_young_dupre} can
then be turned into a differential equation for $x_I(y_I)$, whose
solution reads
\begin{equation}\label{eq:interface_shape}
x_I(y_I)-x_I(L_y/2)=\int_{L_y/2}^{y_I}dy\,\frac{\gamma_{\rm LS}-\gamma_{\rm GS}+F_D(y)-y\Delta p}{\sqrt{\gamma_{\rm LG}^2-[\gamma_{\rm LS}-\gamma_{\rm GS}+F_D(y)-y\Delta p]^2}}\;.
\end{equation}
The predictions of Eq.~\eqref{eq:interface_shape} are shown as red curves in
Fig.~\ref{newfig3}c, and compare well with the interfaces measured in the
simulations. The Active Young-Dupré equation can thus be used to
account for the full interface shapes.

Finally, we stress that partial wetting, leading to a finite contact angle, is only possible thanks to the strong vortex flow created at the triple point (Fig.~\ref{newfig3}b). Indeed, solving Eq.~\eqref{eq:interface_shape} for $F_D=0$ predicts a circular arc of radius $\gamma_{\rm LG}/\Delta p$, which is shown in green in Fig.~\ref{newfig3}c. In all our simulations, this current-free solution systematically fails to describe the interface. Partial wetting is thus only possible in this system thanks to the presence of steady-state currents. 

We note that, contrary to standard active ratchets~\cite{reichhardt2017ratchet}, steady currents are not here a simple consequence of parity symmetry breaking by an external potential since, unlike in~\cite{zakine2020}, the confining walls are invariant by translation along $\hat x$. Instead the currents emerge from a subtle feedback mechanism: They result from the spontaneous symmetry breaking induced by the liquid-gas interface, which is itself stable only thanks to the currents.

Let us now show that, in addition to their role in stabilizing partial wetting, the steady currents also lead to a physics of wetting very different from that of equilibrium systems.

\begin{figure*}
  \centering
  \includegraphics{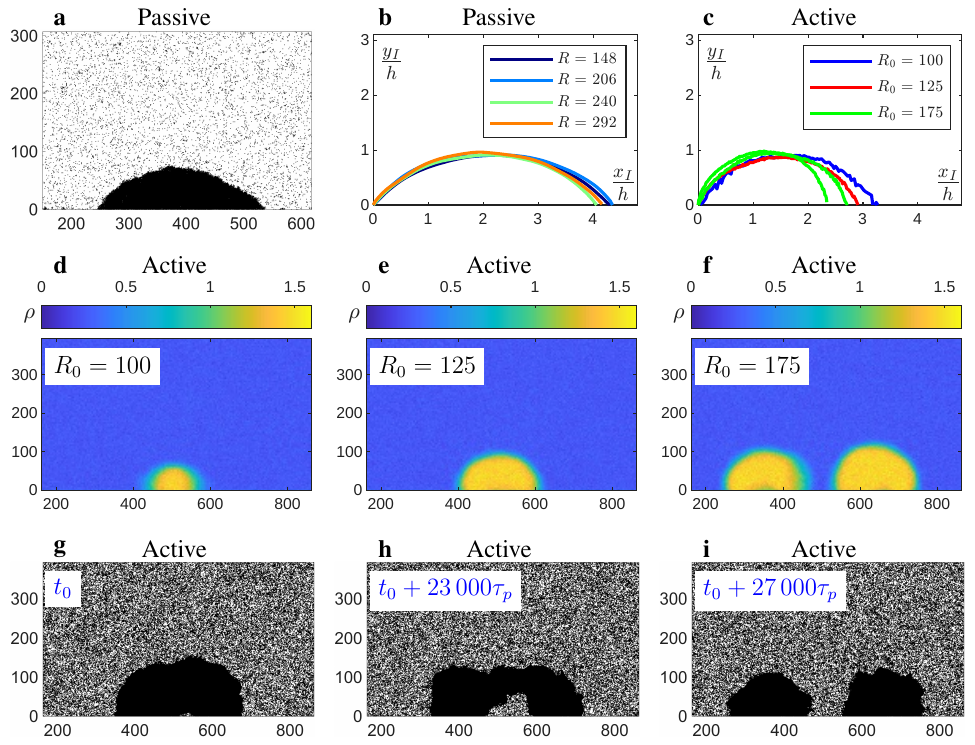}
\caption{{\bf Size selection and splitting dynamics of active droplets adsorbed on a solid surface.} \textbf{a, b}, Passive droplets have a size-independent contact angle. A snapshot is shown in panel a while panel b shows the average interfaces, upon rescaling $x$ and $y$ by the droplets heights. This demonstrates the scale-free nature of the droplet in the passive case. \textbf{c, d, e, f}, Average active-droplet shape as their initial radii $R_0$ increase. Panels d,e,f show the average density profile of the droplets. Large values of $R_0$ lead to the splitting of the initial droplet. Panel c shows the lack of scale-free nature of the active system: the current field generated by the droplet is size-dependent, hence selecting acceptable values of $(\varphi,R)$. \textbf{g, h, i}, Unlike in the passive case, large droplets are unstable and split, leading to a rich dynamics illustrated in these successive snapshots and in SM Movie~2.}
  \label{newfig4}
\end{figure*}

\section{Droplet instability and intermittent dynamics}

In equilibrium, the solution of the passive Young-Dupré
equation---including its correction due to Laplace pressure---is a
circular segment with a fixed contact angle such that
$\cos\varphi_{\rm eq}=\frac{\gamma_{\rm GS}-\gamma_{\rm
    LS}}{\gamma_{\rm LG}}$ (Fig.~\ref{newfig4}a). Increasing the average
density then leads to droplets of increasing radii at fixed contact
angles $\varphi_{\rm eq}$, that can all be collapsed onto a master
curve by dividing the interface coordinate $(x_I,y_I)$ by the droplet
height $h$: The equilibrium system admits scale-free
solutions illustrated in Fig.~\ref{newfig4}b.

In the active case, the active Young-Dupré equation also includes the
drag force, and the Laplace pressure is dressed by currents (See
SM Section~\ref{SM:sec:LaplacePressure}). Since the current field is a complex, non-local functional of the full system configuration, droplets of increasing sizes lead to different values of $F_D$, so that the droplet interface cannot be scale free. In particular, a macroscopic droplet
requires a diverging radius of curvature, leading to a locally
straight interface, as depicted in Fig.~\ref{newfig1}a. This requires
an angle $\phi(y_b)$ independent of $y_b$, which is in direct
contradiction with the fact that $F_D(y_b)$ varies with $y_b$. As
such, we expect the active Young-Dupré equation to forbid macroscopic
active droplets.

To test this prediction, we simulated droplets of increasing sizes and
report their average density profiles and interfaces on
Figs.~\ref{newfig4}d-f and Fig.~\ref{newfig4}c, respectively. As the
density increases, not only do the average profiles vary, but large
droplets break into smaller ones (Fig.~\ref{newfig4}f). First, the drag
force selects admissible droplet sizes, hence breaking the scale-free
nature of the passive system. Then, like any object embedded in an
active fluid (see~\cite{granek2023inclusions} for a review), the droplets have complex interactions mediated by the particle currents. The result is a rich intermittent dynamics with many merging and splitting
events illustrated in~Fig.~\ref{newfig4}g-i and in SM Movie~2. 
Note that, for the same parameters,  Fig.~\ref{newfig2}a shows that phase-separated profiles are stable over time in the absence of confining walls. (See also SM Section~\ref{SM:sec:stability} and SM Movies~3 and 4.) The droplet instability reported here thus has nothing to do with well established bulk instabilities leading to microphase separation~\cite{tjhung2018cluster,caballero2022activity}.

For larger persistence lengths, we observe a second scenario leading to droplet size selection, distinct from the growing-and-splitting mechanism reported in Fig.~\ref{newfig4}. Indeed, SM Fig.~\ref{SM:fig:reverse-ostwald} and Movie 5 show a reversed Ostwald-ripening phenomenon leading to the growth of small distant droplets at the expense of the large initial droplet. Again, this is a surface effect since systems of comparable sizes show a standard Ostwald ripening when the confining walls are replaced by periodic boundary conditions, leading to complete phase separation. 

All in all, our results thus show that the physics of adsorbed active droplets is very rich. The active Young-Dupré equation that we have derived here captures their mechanical equilibria. Together with recent progress in predicting the emergent currents in active fluids~\cite{granek2023inclusions}, our results thus pave the way to a complete theory for the dynamics and instability mechanisms of adsorbed droplets.

\section{Conclusion.}

In this article, we have laid out the foundations for a theory of capillary phenomena in active matter, from the determination of forces exerted along interfaces to the physics of active wetting and adsorbed droplets.

First, we have provided an unambiguous first-principles derivation of the liquid-gas surface tension $\gamma_{\rm LG}$ of an active fluid undergoing motility-induced phase separation. Our results show that the liquid-gas interfaces push on the confining wall, in contrast to equilibrium fluids: the mechanical liquid-gas surface tension is indeed negative. An important prediction that can be tested experimentally is the expulsion of partially-immersed objects from the liquid phase, which, again, is at odds with equilibrium phenomenology.

Second, deriving an Active Young-Dupré equation, we have identified the key ingredient that allows partial wetting in active fluids undergoing motility-induced phase separation.
The breaking of parity symmetry in the vicinity of the triple point leads to the emergence of large vortices of steady-state particle currents. The resulting steady drag stabilizes the imbalance between the mechanical surface tensions at play, leading to a finite contact angle at the triple point. These steady currents lead to new physics for adsorbed droplets: their large-scale structure and their non-local dependence on the system configuration prevent the existence of scale-free droplet solutions, leading instead to a rich dynamics with intermittent merging and splitting events.

Our results should first be put to work in the context of synthetic active colloids~\cite{palacci2010sedimentation,buttinoni2013dynamical}, for which quantitative contacts between microscopic active models and experiments have already been achieved~\cite{ginot2015nonequilibrium,ginot2018sedimentation,lefranc2025synthetic}. Recent results on the partial wetting of solid boundaries by active fluids appear as an ideal test ground for our theory~\cite{fins2024steer}. Partial wetting is also observed in many biological active systems~\cite{douezan2011spreading,perez2019active,agudo2021wetting,adkins2022dynamics,tayar2023controlling,pallares2023stiffness}. 
There, in addition to the steady drag, our Active Young-Dupré equation will also need to comprise terms to account for possible momentum sources that would violate the equations of state for pressure and surface tension~\cite{solon2015natphys,junot2017active}, or for the coupling to momentum-conserving fluids. Already, the physics that we uncovered will have to be considered when studying the wetting of surfaces by cellular tissues~\cite{pallares2023stiffness}.

\section{Acknowledgments}
FvW, JT \& RZ thank K. Mandadapu at UC Berkeley for hospitality and
discussions, JT thanks the Laboratoire MSC for hospitality. YZ acknowledges support from National Natural Science
Foundation of China (Grant 12304252). YK acknowledges
  financial support from ISF (2038/21) and NSF/BSF (2022605). FvW and JT acknowledge support from ANR Thema.

\section{Author Contributions}
YZ, RZ, AD, JT, YK and FvW conceived the project. YZ, AD, JT, YK and FvW wrote the manuscript. YZ produced all numerical data.

\section{Competing Interests}
The authors declare no competing interests.

~\newpage{}

\section{Methods}\label{sec:methods}

\subsection{Simulations of active and passive systems.} Here we describe the models used for active and passive simulations, the modeling of the particle-wall interactions, and the numerical implementations of all models.

\noindent\textit{Active dynamics.}  
Our complete active dynamics read
\begin{align}
    \dot{\rr}_i&=v_0\uu(\theta_i)-\mu\sum_{j\neq i}\nabla_{\rr_i}U(\rr_i-\rr_j)-\mu\nabla_{\rr_i}U_w(\rr_i)\;, \label{met:eomr}\\
\dot{\theta}_i&=\sqrt{2/\tau_p}\,\xi_i\;.\label{met:eomtheta}
\end{align}
The particle interaction potential is modelled using a stiff harmonic potential:
\begin{equation}
    U(\rr)=\epsilon\left(1-\frac{r}{r_{0}}\right)^2\Theta(r_{0}-r)\;,
\end{equation}
with $\epsilon=50$ and $r_0=1$. $\Theta$ is the Heaviside step function. This choice of interaction potential allows for large time-steps and fast simulations but we checked that our results are robust to the detailed form of the repulsive potentials by also using a WCA potential:
\begin{equation}
    U_{\rm WCA}(\rr)=4\epsilon \left(\frac{\sigma^{12}}{r^{12}}-\frac{\sigma^6}{r^6}\right)\Theta(r_0-r)\;,
\end{equation}
where $r_0=2^{1/6}\sigma$. 

In all our simulations, the self-propulsion speed was set to $v_0=5$. The persistence length $\ell_p=v_0\tau_p$ was varied by changing $\tau_p$ from $20/3$ to $100$. 

\noindent\textit{Passive dynamics.}
For the passive dynamics, we consider $N$ coupled overdamped Langevin equations in $d=2$ space dimensions, given by
\begin{equation}
    \dot{\rr}_i=-\mu\sum_{j\neq i} \nabla_{\rr_i} U(\rr_i-\rr_j)-\mu\nabla_{\rr_i} U_w(\rr_i)+\sqrt{2D}\bm{\eta}_i\;,\label{eqn_langevin}
\end{equation}
where $i,j$ are the particle labels, $D=\mu k_BT$ is the particle diffusivity, and $\bm{\eta}_i$ is a zero-mean Gaussian white noise satisfying $\langle \eta_{i,\alpha}(t)\eta_{j,\beta}(t')\rangle=\delta_{ij}\delta_{\alpha\beta}\delta(t-t')$. As in the active case, we always take $\mu=1$.

The particle interaction potential is modelled using a piece-wise polynomial function for $U(\rr)$, with a short-range repulsion and an attractive tail, which allows us to tune independently the repulsion and attraction strengths:
\begin{equation}\label{eqn_x3potential}
    U(\rr)=\epsilon_{r}\left(1-\frac{r}{r_{0}}\right)^2\Theta(r_{0}-r)-\epsilon_a \frac{r}{r_{0}}\left(3-\frac{r}{r_{0}}\right)^2\Theta(3r_{0}-r)\;.
\end{equation}
We again use $r_0=1$. For all passive simulations, we used $D=7$, $\epsilon_r=100$, $\epsilon_a=1$, which leads to phase separation. 

\noindent\textit{Wall-particle interactions.}
Since we use overdamped Langevin dynamics, we have to resolve explicitly the interactions between particles and walls. To do so, we model them using an explicit potential $U_w$. 
Most of our active simulations rely on a harmonic repulsive potential. For instance, the top confining wall is modelled as
\begin{equation}
U_w(x,y;\epsilon_w)=\epsilon_w (y-L_y)^2\Theta(y-L_y)\;.
\end{equation}

For passive particles, repulsive walls do not lead to wetting and  we thus also include attractive tails in the wall-particle interactions using the same functional form as Eq. \eqref{eqn_x3potential}. This time, the---say---top confining wall is modelled as
\begin{equation*}
U_w(x,y;\epsilon_{wa})=\epsilon_{wr}\left(y-L_y-2r_{w0}\right)^2\Theta(y-L_y-2r_{w0})-\epsilon_{wa}(y-L_y)^2(3r_{w0}-y+L_y)\Theta(y-L_y)\;.\label{eqn_passive_wall_potential}
\end{equation*}
In all the passive simulations, we use $r_{w0}=1$ and $\epsilon_{wr}=100$, and $\epsilon_{wa}=5$, which leads to partial wetting.

When needed, the bottom and side walls are built using the same confining potentials. 

\noindent\textit{Numerical implementations.}
To simulate the active dynamics, we use Euler time discretization with time step $\delta t=10^{-3}$ to integrate Eqs. \eqref{met:eomr}-\eqref{met:eomtheta}. To simulate the passive dynamics, we use Heun's method~\cite{saito1993} with time step $\delta t=10^{-4}$ to integrate Eq.~\eqref{eqn_langevin}. 

In both cases, the forces between the particles are calculated using cell lists: The system is binned into cells of linear size equal to the interaction range ($r_0$ for active particles, $3r_0$ for passive particles). The forces on the particles in a cell are then calculated by considering possible interactions with the particles in the same cell and in the $8$ neighbouring cells. 

Due to the heavy numerical requirements of the project, it was necessary to use a C++ code parallelized using message passing interface (MPI) to allow for large-scale long-time simulations. To enable parallel computing, the simulation box is thus divided into blocks governed by different threads. The positions of the particles in the cells at the block boundaries are communicated to the neighbouring threads after every time step. To give a feeling of the required numerical effort, the simulations appearing in Fig.~2c-d and Fig. 3 comprise roughly $6\times 10^5$ particles. These were run using 32 cores of AMD EPYC Zen 3 CPUs with maximal frequency of 3.5GHz and each simulation took around 60 days. The total CPU (wall) time spent in Fig.~2 and Fig.~3 is estimated around $10^2$ CPU years and the overall numerical effort to produce the figures appearing in the main text and the SI is estimated to 10 centuries of CPU time. 

Finally, we note that active matter has been doomed with strong finite-size effects since the birth of the field 30 years ago. Typically, boundary layers and other microscopic structures survive up to scales comparable to the persistence length, which both leads to rich physics~\cite{angelani2011effective,ray2014casimir,kjeldbjerg2021theory} as well as make, e.g., the order of phase transitions very hard to assess~\cite{gregoire2004onset,martin2021fluctuation}. Wetting is no exception to this rule, which explains the numerical effort required to produce the figures included in this article. The reason is that the values of the state variables like pressure and surface tension have to be measured in bulk phases, at macroscopic distances from the boundaries (See SM Section~\ref{SM:sec:finitesize} for a discussion of the finite-size effects).

\subsection{Measurement of the stress-tensor field (Figures~2 \& 3).}

Most measurements that appear in this article are based on the spatial integration of Eq.~\eqref{eq:ss_flow}. As a result, the data that we need to record are also spatial integrals. In practice, we used the cell list to compute these integrals, as well as to produce the heat maps that are shown in the figures. 

\begin{figure}[hb]
    \centering
  \includegraphics{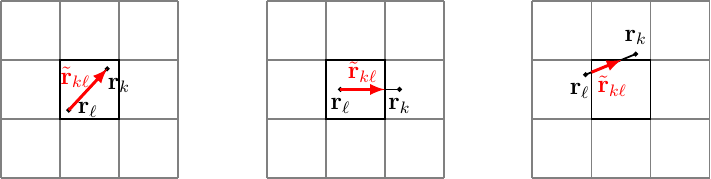}
    \caption{The definition of $\tilde{\rr}_{k\ell}$ in different scenarios. The squares represent cell $(i,j)$.}
    \label{fig:cells}
\end{figure}

We store the integral of the stress tensor over the $(i,j)$-th cell, centered on $(x_i,y_j)=((i+1/2)r_0,(j+1/2)r_0)$, which is computed as
\begin{equation}
  \sigma(x_i,y_j) =  \frac{1}{r_0^2}\int_{ir_0}^{(i+1)r_0}dx\int_{jr_0}^{(j+1)r_0}dy\, \sigma(x,y)\;.\label{eq:meassigma}
\end{equation}
In turn, the various contributions entering the stress tensor are computed as:
\begin{align}
    \rho(x_i,y_j)&=\frac{1}{r_0^2}\left\langle\sum_{k\sim (i,j)} 1\right\rangle\;, \\
    I^{(1)}_{\alpha\beta}(x_i,y_j)+I^{\rm ext}_{\alpha\beta}(x_i,y_j)&=\frac{1}{r_0^2}\left\langle\sum_{k\sim (i,j)} f_{k,\beta} u_{k,\alpha}\right\rangle\;, \\
    Q_{\alpha\beta}(x_i,y_j)&=\frac{1}{r_0^2}\left\langle\sum_{k\sim (i,j)} \left(u_{k,\alpha}u_{k,\beta}-\frac{1}{2}\delta_{\alpha\beta}\right)\right\rangle\;,
\end{align}
where $f_{k,\alpha}=-\sum_l\partial_\alpha U(\rr_k-\rr_l)-\partial_\alpha U_w(\rr_k)$ is the total force experienced by particle $k$ and $k\sim(i,j)$ denotes a summation over particles in the $(i,j)$-th cell. The integration of the IK tensor over cell $(i,j)$ requires a summation over pairs of interacting particles. This includes both particles located in cell $(i,j)$ as well as particles located at $\rr_k$ and $\rr_\ell$ such that $\rr_{k\ell}\equiv \rr_k-\rr_\ell$ has a finite intersection with cell $(i,j)$. The summation is then carried out as~\cite{smith2017towards}:
\begin{equation}
    \sigma^{\rm IK}_{\alpha\beta}(x_i,y_j)=\frac{1}{2r_0^2}\sum_{k \vee \ell}\langle\tilde{r}_{k\ell,\alpha}\partial_\beta U(\rr_k-\rr_\ell)\rangle \;,\label{eq:measIK}
\end{equation}
where $\tilde{\rr}_{k\ell}$ is the fraction of $\rr_{k\ell}$ that lies inside the $(i,j)$-th cell and $k \vee \ell$ refers to pairs such that the line segment going from $\rr_k$ to $\rr_\ell$ has a non-zero intersection with the cell $(i,j)$ (see Fig.~\ref{fig:cells}). 

The time average of the observables defined in Eqs.~\eqref{eq:meassigma}-\eqref{eq:measIK} is then calculated by averaging over repeated measurements carried out every two time units. Note that our cell list includes extra space at $x,y\leq 0$ and $x,y\geq L_x,L_y$ to properly measure the full stress-tensor field.

\subsection{Measurement of the fluid-solid surface tensions (Figure 3).}

We measure the fluid-solid surface tensions as~\cite{zakine2020}:
\begin{equation}
    \gamma_{\rm LS,GS}=\int_{-\infty}^{y_b}\sigma_{xx}(x_{\rm L,G},y)\,dy +(y_b-y_w)p_{\rm{L,G}}\;,
\end{equation}
where $y_b$ is an ordinate deep in the bulk of the liquid and gas phases, such that $y_b\gg \ell_p$. $x_{\rm L,G}$ are abscissa deep in the liquid and gas phases. In practice, we take $y_b=100$ and average $\gamma_{\rm LS, GS}$ over the top and bottom walls. We take $x_{\rm G}=L_x/2$ and $x_{\rm L}=0$.

\subsection{Droplet setups (Figures 1 and 4).} 

To simulate the active droplets shown in Fig.~1c and Fig.~4 of the main text, we confine the system along the $y$-axis using walls located at $0$ and $L_y$. We use periodic boundary conditions along the $x$-axis. We initialize the simulations with semi-circular liquid droplets of radii $R_0$ in contact with the wall and complement the system with a gas phase. The orientations of the particles are random. The gas and liquid densities are taken from the measurement of the binodal densities $\rho_{\rm L}$ and $\rho_{\rm G}$, measured using independent simulations with periodic boundary conditions. Note that because of the wetting of the walls by finite boundary layers, some particles initially located inside the droplet will leak outside as the system relaxes to steady state.

In Fig.~\ref{newfig1}c, for the active droplet, we use $v_0=5$, $\tau_p=20$, $L_x=1536$, $L_y=685$, $\epsilon_w=0.08$ for the bottom wall and $\epsilon_w=0.05$ for the upper wall. (We use a softer top wall to minimise the liquid condensation on the top wall.) The initial radius of the droplet is $R_0=250$, and we show a snapshot at $t=4.5\times 10^4$ time units.
In Fig.~\ref{newfig4}, we use $v_0=5$, $\tau_p=10$, $L_x=1024$, $L_y=450$, $\epsilon_w=0.1$ for the bottom wall and $\epsilon_w=0.08$ for the upper wall. The systems are then simulated for a duration of $3\times 10^4\tau_p$. The density fields in Fig.~\ref{newfig4}d-f are averaged over a time window $t\in [2.7,2.9]\times 10^4\tau_p$ in the center-of-mass frame. Note that the active droplets have complex dynamics. The density fields shown in Fig.~\ref{newfig4}d-f thus correspond to mesoscopic time averages, long enough to get a well-defined average shape for the droplet, but short enough that merging and splitting events do not occur on this timescale. The liquid-gas interfaces are then identified as the positions where the density equals $(\rho_{\rm G}+\rho_{\rm L})/2$. Fig.~\ref{newfig4}c shows the aligned rescaled interfaces. For the case with two droplets shown  in Fig.~\ref{newfig4}f, the interfaces are averaged separately and, for symmetry reasons, aligned using their outward angle so that the interface of the right-most droplet shown in panel Fig.~\ref{newfig4}f is mirrored along $x$ when plotted in Fig.~\ref{newfig4}c.

For passive systems, we use $L_x=768$, $L_y=378$, $\epsilon_{wa}=5$ for the bottom wall and $\epsilon_{wa}=0$ for the top wall. We first let the system equilibrate for $2\times 10^4$ time units and then average the density field for another $2\times 10^4$ time units. The liquid-gas interfaces are identified as the positions where the density equals $(\rho_{\rm G}+\rho_{\rm L})/2$, and the radius $R$ of a droplet is measured by fitting its interface with a circle. 

\subsection{Figures 2 \& 3.} In Fig.~\ref{newfig2}a, $\tau_p=20$, $\epsilon_w=0.2$, $L_x=1206$, $L_y=554$. In Fig.~\ref{newfig3}, $\tau_p=20$, $\epsilon_w=0.1$, $L_x=1206$, $L_y=550$. We detail here the measurements of $\gamma_{\rm LG}$ carried out  using either the bulk setup or the force measurement and Eq.~\eqref{eq:Ftot2}.

\noindent\textit{Bulk measurement of $\gamma_{\rm LG}$.}
To measure $\gamma_{\rm LG}$ from the bulk definition Eq.~\eqref{eq:gamma_LG}, we use a periodic system without any external confinement. We initialize the system in a phase-separated configuration, using the gas binodal density for $x\in [300,906]$ and the liquid binodal density for $x \in [0,300] \cup [906,1206]$. The system then evolves for $5\times 10^4$ time units (at least $500\tau_p$), and we then average the data over the next $2.5\times 10^5$ time units. The density  and stress tensor fields are measured as discussed above and averaged over $y$.

Using the density profile averaged over $y$, we identify  the liquid-gas interface as the abscissa $x_I$ where the density equals $(\rho_{\rm G}+\rho_{\rm L})/2$. (The precise location is found by interpolating the density measured in the simulation cells using a linear function in the vicinity of $x_I$.) Then we integrate $\sigma_{yy}-\sigma_{xx}$ over $[x_I-\ell, x_I+\ell]$, where $\ell$ is chosen sufficiently large that the stress tensor is isotropic at $x_I\pm \ell$. Specifically, we use $\ell=\{75,75,75,75,200,300\}$ for $\tau_p=\{20/3,10,12.5,20,50,100\}$, respectively. 

\noindent\textit{Force measurement of $\gamma_{\rm LG}$ through Eq.~\eqref{eq:Ftot2}.} We consider the system shown in Fig.~\ref{newfig2}a with confining walls at $y_w=0, L_y$. We used either a  stiff wall ($\epsilon_w=2$) or a soft wall ($\epsilon_w=0.1$). The liquid phase fully wets the stiff wall but it only partially wets the soft wall when $\tau_p\leq 50$. We then initialize the system in a phase-separated configuration as described above. The system is then first evolved over $5\times 10^4$ time units (at least $500\tau_p$), before we average the data over the next $2.5\times 10^5$ time units.

To measure $\gamma_{\rm LG}$ from Eq.~\eqref{eq:Ftot2}, we first compute the average density and stress profiles in the bulk of the system, $\rho_{\rm av}(x)$ and $\sigma_{\rm av}(x)$, by averaging the data over $y\in[L_y/2-50,L_y/2+50]$. We again identify the abscissa of the liquid-gas interfaces $x_{1,2}$ by finding the location where $\rho_{\rm av}(x)=(\rho_{\rm G}+\rho_{\rm L})/2$. The lengths of gas and liquid phases are estimated as $\ell_{\rm G}=|x_2-x_1|$ and $\ell_{\rm L}=L_x-\ell_{\rm G}$, respectively\footnote{Note that, in Eq.~\eqref{eq:Ftot2}, it may seem more natural to define the interface position as the abscissa where $-\sigma_{xx}$ equals $(p_{\rm G}+p_{\rm L})/2$. This leads to negligible differences in the value of $\gamma_{\rm LG}$.}. The bulk pressures $p_{\rm G}$ and $p_{\rm L}$ are calculated by averaging $-\sigma_{{\rm av},xx}$  over $x\in[L_x/2-50,L_x/2+50]$ and $x\in[0,50]\cup[L_x-50,L_x]$, respectively. 

\subsection{Wilhelmy plates (Figures~\ref{newfig2}, \ref{SM:fig:Wilhelmy}, \ref{SM:fig:sigmaxy}, \ref{SM:fig:trsigma}, and~\ref{SM:fig:bulk_vs_wilhelmy}).} 
For the Wilhelmy-plate simulations shown in Figs.~\ref{newfig2}c-d, \ref{SM:fig:Wilhelmy}, \ref{SM:fig:sigmaxy}, \ref{SM:fig:trsigma}, and \ref{SM:fig:bulk_vs_wilhelmy}, we pin the position of the liquid phase by confining the system along the $\hat x$ directions using confining walls ($\epsilon_w=2$). We consider periodic boundaries in the $y$ direction at $y=\pm L_y/2$. We impose left-right parity symmetry in the system to minimize the influence of confining boundaries: we initialize the system with two liquid phases for $x < 296$ and $x > L_x-296$ (we use $L_x=1528$ and $L_y=512$ unless specified otherwise), and we insert a Wilhelmy plate in the vicinity of each liquid-gas interface. Each plate interacts with the particles through a repulsive potential:
\begin{equation}
    U_p(\rr)=u_p(\rr-{\bf r}_{p1})+u_p(\rr-{\bf r}_{p2})\;,
\end{equation}
where ${\bf r}_{p1}$ and ${\bf r}_{p2}$ are the positions of two plates.  We use $u_p(\rr)=\epsilon_p(r-R_p)^2\Theta(R_p-r)$, where $R_p$ is the radius of the plate and $\epsilon_p=2$. 

\noindent\textit{Fig.~\ref{newfig2}d}. We use $R_p=32$ and $M=1$. Initially, the two plates are placed at $\rr_{p1}=(316,0)$ and $\rr_{p2}=(L_x-316,0)$. The system is first relaxed for $9.7\times 10^3$ time units and capillary bridges spontaneously form between the plate and the liquid phase. Then the plates are slowly pulled out of the liquid by changing $\rr_{p1,2}$ at a speed $0.01r_0$ per unit time for $6\times 10^3$ time units, which gives a total distance of $60r_0$ from the initial position. The plates are then held fixed and the system is set to relax for $1.02\times 10^4$ time units. Finally, the plates are released and evolve according to the Hamiltonian dynamics \eqref{eq:Fwilhelmy}. The trajectories shown in Fig.~\ref{newfig2}d correspond to a duration of $2.3\times 10^3$ time units.

\noindent\textit{Figs.~\ref{newfig2}c, \ref{SM:fig:Wilhelmy}, \ref{SM:fig:sigmaxy}, \ref{SM:fig:trsigma}, and \ref{SM:fig:bulk_vs_wilhelmy}}. Initially, the two plates are placed at $\rr_{p1}=(300,0)$ and $\rr_{p2}=(L_x-300,0)$. The system is first relaxed for $10^3$ time units and capillary bridges form between the plates and the liquid phase. Then the plates are slowly pulled out of the liquid by changing $\rr_{p1,2}$ at a speed $0.01r_0$ per unit time for $4\times 10^3$ time units, which gives a total distance of $40r_0$ from the initial position. The plates are then held  fixed and the system is set to relax for $2.5\times 10^4$ time units. Data are then measured for the next $4\times 10^4$ time units. This protocol is repeated on several independent runs and the data are averaged over the two plates in each setup. We note that we only recorded the force exerted on the plate provided a capillary bridge still connects the plate to the liquid phase at the end of the data-averaging sequence. 

To build the various contributions entering panel e of Fig.~\ref{SM:fig:Wilhelmy}, we proceed as follows. The forces exerted on the plates are measured as detailed above. $F_D$ is measured by integrating $J_x$ over the dashed red region in Fig.~\ref{SM:fig:Wilhelmy}a. The bulk pressures $p_{\rm L}$ and $p_{\rm G}$ are measured by averaging $-{\rm Tr}\,\sigma/2$ in the regions $(x,y)\in [110,170]\times[-50,50]$ and $[L_x/2-6,L_x/2]\times[-L_y/2+1,L_y/2-1]$, respectively. To measure the angle $\phi$, we first identify the positions of the interface $x=x_I(y)$ as the abscissa where $\rho(x_I,y)=(\rho_{\rm G}+\rho_{\rm L})/2$. Next, we numerically calculate the derivative $\hat x_I'(y)$ as:
\begin{equation}
    \hat x_I'(y)\simeq \frac{x_I(y+r_0)-x_I(y-r_0)}{2r_0}\;.
\end{equation}
The resulting values of $\hat x_I'(y)$ are noisy and we average them as $x_I'(y)=[\hat x_I'(y) + \hat x_I'(y\pm r_0) + \hat x_I'(y\pm 2r_0)]/5$. We then extract the angle as $\phi=\arctan[ 1/x_I'(y)]$. Finally, we use the symmetry of the system to average $\cos\phi(h/2)$ and $\cos\phi(-h/2)$ to further reduce fluctuations. To measure the surface tension $\gamma_{\rm LG}$, we integrate $\sigma_{yy}-\sigma_{xx}$ over $[x_I-\ell,x_I+\ell]$ with $\ell=75$ at $y=\pm L_y/2$. For consistency, we check in Fig.~\ref{SM:fig:bulk_vs_wilhelmy} that the measured values are consistent with the bulk measurements using the phase-separated setup shown in Fig.~\ref{newfig2}a.

\subsection{Estimation of statistical errors.} Throughout the article,  error bars represent statistical uncertainties, estimated as three times the standard error of the mean of the measured observables. To estimate the standard error of the mean, we use a block-average strategy~\cite{flyvbjerg1989}. First, we sample our observables, denoted $X(t)$ for clarity, at high frequency, every $\Delta t=2$ time units. We then calculate a block average $\bar{X}(t)=\frac{1}{N}\sum_{k=0}^{N-1} X(t+k\Delta t)$ using a block size $N=50$. Since $N\Delta t\geq\tau_p$, it is reasonable to assume that our successive measurements of $\bar X(t)$ are uncorrelated and we indeed checked for all our observables that this was the case. Then denoting $\bar{X}_i=\bar{X}(iN\Delta t)$, we record the accumulated sum of $\bar{X}_i$ and $\bar{X}_i^2$ respectively, and estimate the standard error as
\begin{equation}
e_X\equiv \sigma_{\langle X\rangle}=\frac{1}{\sqrt{M-1}}\sigma_{\bar{X}}\;,
\end{equation}
where $M$ is the number of blocks and $\sigma_{\bar X}$ denotes standard deviation $\sigma^2_{\bar X}=M^{-1}\sum_i \bar X_i^2-M^{-2}(\sum_i \bar X_i)^2$. Provided $N$ is large enough, this estimate is unaffected by the correlations in $X(t)$~\cite{flyvbjerg1989}.

For spatial averages, we assume (and verify numerically) that the fluctuations of the density and stress fields are spatially uncorrelated in homogeneous bulk phases on our coarse-graining scale.

\section{Data availability}
The data that support the findings of this study are available from
the corresponding author upon request. The source data generated in
this study have been deposited in the Figshare database under
accession address \href{https://doi.org/10.6084/m9.figshare.31148908}{https://doi.org/10.6084/m9.figshare.31148908}.

\section{Code availability}
The codes and algorithms that have been used to support the findings
of this study are available from the corresponding author upon
request.

\clearpage
\addtocontents{toc}{\protect\setcounter{tocdepth}{2}}
\setcounter{section}{0}
\setcounter{figure}{0}
\setcounter{equation}{0}
\renewcommand{\thefigure}{S\arabic{figure}}
\renewcommand{\theequation}{S\arabic{equation}}
\renewcommand{\thesection}{S\arabic{section}}
\renewcommand{\theHfigure}{S\arabic{figure}}
\renewcommand{\theHequation}{S\arabic{equation}}
\renewcommand{\theHsection}{S\arabic{section}}
\captionsetup[figure]{font=normalsize,labelfont={bf},labelformat={default},labelsep=period}
\spacing{1.1}
\setlength{\parskip}{9pt}

{\noindent\Large\bfseries\textsf{Supplementary Material for ``The Active Young-Dupr\'e Equation''}\par}
\tableofcontents{}

Throughout the document, references to Figures and equations of the main text are done as ``Fig.~n" and ``Eq.~(n)", whereas supplementary figures and equations are referred to as ``Fig.~Sn" and ``Eq.~(Sn)".

\section{Descriptions of supplementary movies}

\begin{itemize}
  \item SM movie 1: Wilhelmy plates being expelled from the active liquid phase. Blue dots represent particle positions and red lines show the position of the confining walls. The system size is given by $L_x=1528$ and $L_y=512$. Initially, the two plates are positioned at $\rr_{p1}=(316,0)$ and $\rr_{p2}=(L_x-316,0)$ and the system relaxes for $9.7\times 10^3$ time units. Then the plates are slowly pulled out of the liquid at a speed of $0.01r_0$ per unit time for a duration of $6\times 10^3$ time units, leading to a total displacement of $60r_0$. The plates are then held fixed and the system relaxes for $1.02\times 10^4$ time units. Finally, the plates are released and move according to Hamiltonian dynamics. The movie corresponds to the parameters of Fig.~\ref{newfig2}d: $v=5$, $\tau_p=20/3$, plate mass $M=1$, and plate radius $R_p=32$. 
  \item SM movie 2: Intermittent dynamics of active droplets. The initial radius of the droplet is $R_0=175$. The blue dots represent the positions of the particles, and the red lines show the positions of the confining walls. The movie corresponds to the parameters of Fig.~\ref{newfig4}f-i: $v=5$, $\tau_p=10$,  $\epsilon=50$, $\epsilon_{w,\rm bottom}=0.1$, $\epsilon_{w,\rm top}=0.08$, $L_x=1024$, and $L_y=450$.
  \item SM movie 3: A droplet of initial radius $R_0=175$ with $\ell_p=50$ and $\epsilon=50$ in a periodic system, extending up to $3\times 10^5\tau_p$, showing the stability of macroscopic phase separation. The blue dots represent the positions of the particles. Parameters: $v=5$, $\tau_p=10$, $L_x=L_y=512$.
  \item SM movie 4: A droplet of initial radius $R_0=1750$ with $\ell_p=50$ and $\epsilon=50$ in a periodic system, extending up to $3\times 10^4\tau_p$, showing the stability of macroscopic phase separation. The blue dots represent the positions of the particles. Parameters: $v=5$, $\tau_p=10$, $L_x=L_y=5000$.
  \item SM movie 5: Intermittent dynamics of adsorbed active droplets showing a reverse Ostwald ripening. The gray scale shows the time-lapse density fields and the arrows represent the particle currents. The lengths of the arrows are proportional to $\log|\J/10^{-6}|$. Colors encode the current direction. Parameters: $v=5$, $\tau_p=20$, $\epsilon=52$, $L_x=1024$, $L_y=432$.
  \item SM movie 6: A system with bulk phase separation with $\ell_p=100$ and $\epsilon=55$ in a periodic system. The initial length of the liquid bulk is 512. The blue dots represent the positions of the particles. Parameters: $v=5$, $\tau_p=20$, $L_x=2L_y=1024$.
  \item SM movie 7: Intermittent dynamics of active droplets corresponding to Fig.~\ref{SM:fig:reverse-ostwald}h. The initial radius of the droplet is $R_0=350$. The gray scale shows the time-lapse density fields, and the arrows represent the particle currents. The lengths of the arrows are proportional to $\log|\J/10^{-6}|$. Colors encode the current direction. Parameters: $v=5$, $\tau_p=20$, $\epsilon=100$, $\epsilon_{w,\rm bottom}=0.1$, $\epsilon_{w,\rm top}=0.05$, $L_x=1536$, and $L_y=688$.
\end{itemize}

\section{Time-evolution of the density field  and generalized stress tensor}\label{SM:sec:stress}

To make this manuscript self-contained and to use consistent notations, we follow \cite{solon2018generalized} and construct explicitly the dynamics of the density field of active particles interacting via pairwise forces that are given in Eq.~\eqref{eq:eomr}. We first introduce the density, orientation and nematic-order fluctuating fields, which are defined as:
\begin{align}
    &\hat{\rho}(\rr,t)=\sum_i\delta(\rr-\rr_i(t))\;,\ \hat{\mm}(\rr,t)=\sum_i\uu_i\delta(\rr-\rr_i(t))\;, \\
    &\hat{\Q}(\rr,t)=\sum_i\left[\uu_i(t)\otimes\uu_i(t)-\frac{1}{d}\mathbb{I}\right]\delta(\rr-\rr_i(t))\;,
\end{align}
where $\mathbb{I}$ is the identity matrix, and $d=2$ is the space dimension. Their statistical averages are denoted by $\rho$, $\mm$, and $\Q$, respectively. 

We start from the evolution equation for the single-particle density function $\Psi(\rr,\theta,t)=\langle \sum_i \delta(\rr-\rr_i(t))\delta(\theta-\theta_i(t))\rangle$, which measures the average density of particles at position $\rr$ with orientation $\theta$ at time $t$:
\begin{eqnarray}
\pdiff{\Psi}{t}&=&\nabla_{\rr}\cdot\left(-v_0\uu(\theta)\Psi+\mu\int d\rr' d\theta'\,\Psi^{(2)}(\rr,\theta,\notag\rr',\theta',t)\nabla_{\rr}U(\rr-\rr')+\mu\Psi\nabla_{\rr}U_w(\rr)\right)\\&&+D\Delta_{\rr}\Psi+\tau_p^{-1}\partial_{\theta\theta}\Psi\;. \label{SM:eqn_FK_1}
\end{eqnarray}
We include translational diffusion $D=\mu k_B T$ for generality. The equation contains the two-particle density function $\Psi^{(2)}(\rr,\theta,\rr',\theta',t)$:
\begin{equation}
\Psi^{(2)}(\rr,\theta,\rr',\theta',t)=\left\langle\sum_{i=1}^N\sum_{k\neq i}\delta(\rr-\rr_i(t))\delta(\theta-\theta_i(t))\delta(\rr'-\rr_k(t))\delta(\theta'-\theta_k(t))\right\rangle
\end{equation}
and it is thus not closed. The dynamics of $\Psi^{(2)}$ can be calculated in a similar fashion, and it involves the three-particle density function $\Psi^{(3)}(\rr,\theta,\rr',\theta',\rr'',\theta'',t)$. This procedure can be carried out sequentially to obtain an infinite series of equations, similar to the BBGKY hierarchy for equilibrium systems. 

Next we use Eq.~\eqref{SM:eqn_FK_1} to calculate the dynamics of the average density field $\rho(\rr,t)$. For the sake of clarity, we  use $\partial_\alpha\equiv\partial/\partial r_\alpha$ and always sum over repeated indices. The evolution equation of $\rho$ can be calculated by integrating Eq.~\eqref{SM:eqn_FK_1} over $\theta$ to get
\begin{equation}\label{SM:eqn_continuity}
\pdiff{\rho}{t}=-\partial_\alpha J_\alpha\;.
\end{equation}
The dynamics Eq.~\eqref{SM:eqn_continuity} has the form of a continuity equation
where the current ${\bf J}$ is defined as
\begin{align}
J_\alpha=&v_0m_\alpha-\mu\int d\rr'\,\rho^{(2)}(\rr,\rr',t)\partial_\alpha U(\rr-\rr')-\mu\rho\partial_\alpha U_w(\rr)- D\partial_{\alpha}\rho \label{SM:eqn_J_two_points} \\
=&v_0m_\alpha+\mu\partial_\beta\sigma^{\rm IK}_{\alpha\beta}-\mu\rho\partial_\alpha U_w(\rr)-D\partial_{\alpha}\rho\;.  \label{SM:eqn_J}
\end{align}
In Eq. \eqref{SM:eqn_J_two_points}, we have defined the two-particle density
\begin{equation}
\rho^{(2)}(\rr,\rr',t)=\left\langle\sum_{i=1}^N\sum_{k\neq i}\delta(\rr-\rr_i(t))\delta(\rr'-\rr_k(t))\right\rangle\;,\label{SM:eqn_density_two_points}
\end{equation}
and $\sigma^{\rm IK}$ in Eq. \eqref{SM:eqn_J} is the Irving-Kirkwood stress tensor\cite{solon2018generalized,irving1950statistical}
\begin{equation}
\sigma^{\mathrm{IK}}_{\alpha\beta}=\frac{1}{2}\int d^2\rr'\,\frac{r'_\alpha r'_\beta}{r'}\frac{dU(r')}{dr'}\int_0^1 d\lambda\,\rho^{(2)}(\rr+(1-\lambda)\rr',\rr-\lambda\rr',t)\;.\label{SM:eqn_IK_stress}
\end{equation}
Furthermore, from Eq.~\eqref{SM:eqn_FK_1} we determine the dynamics of $m_\alpha$ as,
\begin{align}
\pdiff{m_\alpha}{t}=&\partial_\beta\left(-v_0Q_{\alpha\beta}+\mu\int d\rr'\,\partial_{\beta}U(\rr-\rr')\left\langle \sum_{i=1}^N\sum_{k\neq i}u_\alpha(\theta_i)\delta(\rr-\rr_i(t))\delta(\rr'-\rr_k(t))\right\rangle\right. \nonumber \\
&+\mu m_\alpha\partial_\beta U_{w}+D\partial_\beta m_\alpha\Biggr)-\partial_\alpha\left(\frac{v_0\rho}{d}\right)-\frac{m_\alpha}{\tau_p}\;.\label{SM:eqn_m_dynamic1}
\end{align}
We note that  Eq.~\eqref{SM:eqn_m_dynamic1} can be rewritten as
\begin{align}
\pdiff{m_\alpha}{t}=&\partial_\beta\left(-v_0Q_{\alpha\beta}-\mu I^{(1)}_{\alpha\beta}-\mu I^{\rm ext}_{\alpha\beta}+D\partial_\beta m_\alpha-\frac{v_0\rho}{d}\delta_{\alpha\beta}\right)-\frac{m_\alpha}{\tau_p}\;,\label{SM:eqn_m_dynamic}
\end{align}
where $\delta_{\alpha\beta}$ is the Kronecker delta, and we define for clarity:
\begin{align}
I^{(1)}_{\alpha\beta}=&-\int d\rr'\,\partial_{\beta}U(\rr-\rr')\left\langle \sum_{i=1}^N\sum_{k\neq i}u_\alpha(\theta_i)\delta(\rr-\rr_i(t))\delta(\rr'-\rr_k(t))\right\rangle\;, \\
I^{\rm ext}_{\alpha\beta}=&-m_\alpha\partial_\beta U_{w}\;.
\end{align}

In the steady state, we have that $\partial_t\rho=\partial_t m_\alpha=0$, leading to $\partial_\alpha J_\alpha=0$ so that $J_\alpha(\rr,t)$ is divergence-free, and Eq.~\eqref{SM:eqn_m_dynamic} leads to
\begin{equation}
m_\alpha=\tau_p\partial_\beta\left(-v_0Q_{\alpha\beta}-\mu I^{(1)}_{\alpha\beta}-\mu I^{\rm ext}_{\alpha\beta}+D\partial_\beta m_\alpha-\frac{v_0\rho}{d}\delta_{\alpha\beta}\right)\;.
\end{equation}
Substituting into Eq.~\eqref{SM:eqn_J}, we find that, in the steady-state:
\begin{align}
\frac{1}{\mu}J_\alpha=&\partial_\beta\left[-\frac{\left(D_{\rm eff}+D\right)}{\mu}\rho\delta_{\alpha\beta}-\frac{dD_{\rm eff}}{\mu} Q_{\alpha\beta}-\ell_p I^{(1)}_{\alpha\beta}-\ell_p I^{\rm ext}_{\alpha\beta}+\frac{D}{\mu}\ell_p\partial_\beta m_\alpha+\sigma^{\rm IK}_{\alpha\beta}\right]-\rho\partial_\alpha U_w(\rr) \nonumber \\
=&\partial_\beta\sigma_{\alpha\beta}-\rho\partial_\alpha U_w(\rr)\;,\label{SM:eqn_force_balance_app}
\end{align}
where $D_{\rm eff}=v_0^2\tau_p/d$ is the effective diffusivity of free active particles, and $\ell_p=v_0\tau_p$ is the persistence length.

Equation~\eqref{SM:eqn_force_balance_app} can be read as a force balance equation, which endows $\sigma$ with the mechanical interpretation of a stress tensor, with
\begin{equation}
\sigma_{\alpha\beta}=-\mu^{-1}\left(D_{\rm eff}+D\right)\rho\delta_{\alpha\beta}-d\mu^{-1}D_{\rm eff} Q_{\alpha\beta}-\ell_p I^{(1)}_{\alpha\beta}-\ell_p I^{\rm ext}_{\alpha\beta}+\mu^{-1}D\ell_p\partial_\beta m_\alpha+\sigma^{\rm IK}_{\alpha\beta}\;.\label{SM:eq:full_stress}
\end{equation}
We note that the stress tensor for passive systems is obtained by setting $\ell_p=0$ and $D_{\rm eff}=0$, which gives back the standard expression
\begin{equation}
\sigma^{\rm passive}_{\alpha\beta}=-\frac{D}{\mu}\rho\delta_{\alpha\beta}+\sigma^{\rm IK}_{\alpha\beta}\;.
\end{equation}

The full stress tensor Eq.~\eqref{SM:eq:full_stress} comprises two contributions $\bm{\sigma}=\bm{\sigma}^a+\bm{\sigma}^{\rm passive}$. Note that, for simplicity, we take $D=0$ when studying active systems (but $D=\mu k_BT\neq 0$ in passive simulations). We note that Eq.~\eqref{SM:eqn_force_balance_app} is exact in the steady state. A physical discussion of the various terms entering $\bm{\sigma}^a$ can be found in~\cite{solon2018generalized}. Throughout the article, we use Eq.~\eqref{SM:eq:full_stress} to carry out measurements of the stress tensor from particle simulations.

\section{Mechanical measurement of the liquid-gas surface tension}\label{SM:sec:LGtension}

In this section we derive Eq.~\eqref{eq:Ftot2}, which relates the definition~\eqref{eq:gamma_LG_speck} of surface tension to an actual
force measurement, starting from the microscopic
dynamics given by Eq.~\eqref{eq:eomr}. To do so, we consider the phase-separated system shown in Fig.~\ref{newfig2}a, which is confined along the $\hat y$ direction by
two walls at the top and bottom of the system and endowed with periodic boundary conditions along $\hat x$. 
The force density on the upper wall can be measured
as $f_w(x)=\int_{y_b}^\infty dy\, \rho(x,y)\partial_{y}U_w(y)$, where
$y_b$ is a position deep in the bulk of the system, far away from the
walls. In the liquid and gas phases, the system exerts an upward force
per unit length on the upper wall equal to the bulk pressures, given by $p_{\rm L}$
and $p_{\rm G}$, respectively\footnote{Note that $p_{\rm G}$ and $p_{\rm L}$ need not be equal due to the curved interface separating the gas and liquid phases.}.

To extract the contributions of the liquid-gas interfaces to the total force  exerted on the upper wall, $F_{w, \rm tot}=\int_0^{L_x}dx' f_w(x')$, we use
Eq.~\eqref{eq:ss_flow}, also given in Eq.~\eqref{SM:eqn_force_balance_app}, to integrate $J_y\equiv \J\cdot\hat y$ over the
upper half of the system, for $y>y_b$ and $0\leq x \leq L_x$. This
leads to
\begin{equation}\label{SM:eq:Ftot1}
F_{w,\rm tot}=   -\int_{0}^{L_x}\!\!\!\!\! dx' \sigma_{yy}(x',y_b)-\int_{y\geq y_b} \!\!\!\!\!\!\!\! d^2 \br \frac{J_y}{\mu}\;,
\end{equation}
where we have used that $\sigma_{xy}=0$ away from the interfaces since
the bulk liquid and gas phases are locally isotropic. In the
steady-state, $\nabla \cdot {\bf J}=0$ implies that the integral of
$J_y$ vanishes. We then note that $-\int_{0}^{L_x}\!  dx'
\sigma_{xx}(x',y_b)=\ell_{\rm G} p_{\rm G} + \ell_{\rm L} p_{\rm L}$, where $\ell_{{\rm G}/{\rm L}}$ are the total lengths of the gas and liquid
phases, respectively. Adding and subtracting this contribution to
Eq.~\eqref{SM:eq:Ftot1} then leads to
\begin{equation}\label{SM:eq:Ftot2}
F_{w,\rm tot} = -n_{{\rm LG}} \gamma_{\rm LG}+ \ell_{\rm G} p_{\rm G} + \ell_{\rm L} p_{\rm L} \;,
\end{equation}
where $n_{\rm LG}$ is the number of liquid-gas
interfaces and we identify the liquid-gas surface tension
$\gamma_{\rm LG}$ as
\begin{equation}\label{SM:eq:gamma_LG}
\gamma_{\rm LG}=\int_{x_{\rm L}}^{x_{\rm G}}dx\,[\sigma_{yy}(x,y_b)-\sigma_{xx}(x,y_b)]\;.
\end{equation}

\section{The Wilhelmy plate experiment}\label{SM:sec:wilhelmy}

\begin{figure}
  \begin{tikzpicture}
    \def\y{4}
    \def\r{.3}    
    \begin{scope}[xshift=-10.5cm]
    \def\xi{-.15}
    \def\yi{.55}
      \path(0.15,\y+2.1) node{\includegraphics{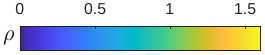}};
      \path(0,\y) node{\includegraphics{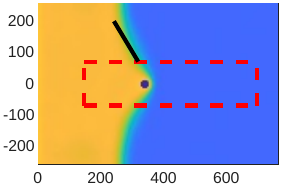}};      
      \draw[ultra thick] (\xi+\r,\y+\yi) arc (0:105:\r);
      \draw  (\xi+\r,\y+\yi)  node[xshift=.1cm,yshift=.3cm] {\large\bf  $\phi$};
      
      \draw[red,<->,ultra thick] (-1.2,\y-.2) -- (-1.2,\y+.6);
      \draw[red, ultra thick] (-1.2,\y+0.2) node[anchor=east] {\large $h$};
      \node at (1.5,\y+0.9) {$\red{\mathcal{A}}$};
      
      \path(0,0) node{\includegraphics{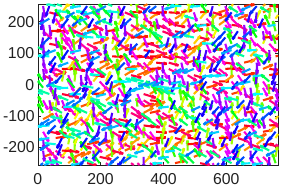}};
      \def\xa{-1.3}
      \def\ya{1.2}
      \def\dw{.25}
      \filldraw[white] (\xa-\dw,\y+\ya-\dw) rectangle (\xa+\dw,\y+\ya+\dw);
      \draw (\xa,\y+\ya) node {\bf (a)};

      \def\xc{-1.3}
      \def\yc{1.2}
      \def\dw{.3}
      \filldraw[white] (\xc-\dw,\yc-\dw) rectangle (\xc+\dw,\yc+\dw);
      \draw (\xc,\yc) node {\bf (c)};
      \def\xcw{1.5}
      \def\ycw{1.0}
      \def\cwbw{0.75}
      \def\cwbh{0.5}
      \filldraw[white] (\xcw-\cwbw,\ycw-\cwbh) rectangle (\xcw+\cwbw,\ycw+\cwbh);
      \path(\xcw,\ycw) node{\includegraphics{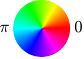}};
    \end{scope}

    \begin{scope}[xshift=-5.5cm]
      \path(0.15,\y+2.1) node{\includegraphics{colorbar_rho_horizontal_wilhelmy}};
      \path(0,\y) node{\includegraphics{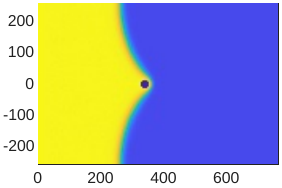}};
      \path(0,0) node{\includegraphics{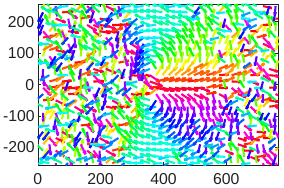}};
      \def\xa{-1.3}
      \def\ya{1.2}
      \def\dw{.25}
      \filldraw[white] (\xa-\dw,\y+\ya-\dw) rectangle (\xa+\dw,\y+\ya+\dw);
      \draw (\xa,\y+\ya) node {\bf (b)};
      \def\xc{-1.3}
      \def\yc{1.2}
      \def\dw{.3}
      \filldraw[white] (\xc-\dw,\yc-\dw) rectangle (\xc+\dw,\yc+\dw);
      \draw (\xc,\yc) node {\bf (d)};
    \end{scope}
    \path (0.5,1.5) node{\includegraphics{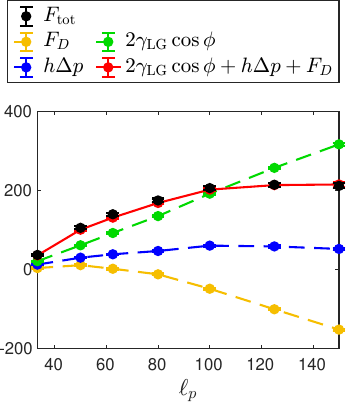}};
    \draw (-1.25,5.45) node {\bf (e)};
  \end{tikzpicture}  
  \caption{Forces exerted on Wilhelmy plates. \textbf{(a-b)} Density fields of active particles with $\tau_p=\frac{20}{3}$ (a) and $\tau_p=20$ (b), speed $v_0=5$ and plate radius $R_p=16$. \textbf{(c-d)} The current fields corresponding to (a) and (b), respectively. The arrow lengths are proportional to $\log|\J/10^{-6}|$ and their colors encode the current direction. \textbf{(e)} The various contributions to the total force $F_{\rm tot}$ on the plate, as a function of $\ell_p$. Error bars represent three times the standard error of the mean (s.e.m.), estimated from $20$ independent simulations, each simulation contains $M=400$ block samples (see \hyperref[sec:methods]{Methods}), giving in total $8000$ samples.}
    \label{SM:fig:Wilhelmy}
\end{figure}

In this section, we calculate the force exerted on a fixed Wilhelmy plate connected to liquids by a capillary bridge. To connect the force exerted on the probe to the liquid-gas surface tension, we integrate the particle current from Eq.~\eqref{eq:ss_flow}, also given in Eq.~\eqref{SM:eqn_force_balance_app}, over a region
$\mathcal{A}=[x_{\rm L},x_{\rm G}]\times [-h/2,h/2]$ that encompasses the disk,
with $x_{\rm G,L}$ abscissa deep in the gas and liquid phases, respectively
(See Fig.~\ref{SM:fig:Wilhelmy}a). This leads to
\begin{equation}\label{SM:eq:Fwilhelmy}
    F_{\rm tot}=\int_{\mathcal{A}} (\partial_x \sigma_{xx}+\partial_y\sigma_{xy})d^2\br+F_D\;,
\end{equation}
where $F_{\rm tot}$ is the average total force exerted by the fluid on
the disk along $\hat x$ and
\begin{equation}\label{SM:eq:drag}
    F_D= -\int_{\mathcal{A}}\frac{J_x}{\mu} d^2\br=\sum_{i| \rr_i\in\mathcal{A}} -\frac 1 \mu \langle\dot x_i\rangle
\end{equation}
is the total drag exerted on the particles contained in the
region $\mathcal{A}$. Since $\sigma_{xy}$ vanishes in bulk phases, the integral of $\partial_y \sigma_{xy}$ in Eq.~\eqref{SM:eq:Fwilhelmy} picks a contribution only from
the interface. The latter is derived in Eq.~\eqref{SM:eq:gammacosphi} in the next subsection, which shows the integral to yield a contribution $\gamma_{\rm LG} \cos\phi$, where
$\phi$ is the angle between the interface and the $\hat x$ axis
(see Fig.~\ref{SM:fig:Wilhelmy}a). All in all, the average total force exerted on the probe is given by
\begin{equation}\label{SM:eq:Fwilhelmy2}
    F_{\rm tot}=2 \gamma_{\rm LG} \cos \phi +h \Delta p +F_D\;,
\end{equation}
where $\Delta p=p_{\rm L}-p_{\rm G}$ is the pressure jump between the liquid and
gas phases due to the curved interface~\cite{solon2018generalized}.
Independent measurements of the contributions entering
Eq.~\eqref{SM:eq:Fwilhelmy2} are shown in Fig.~\ref{SM:fig:Wilhelmy}e, for a
plate held fixed, together with maps of the average density and
current fields for two different persistence lengths. These
measurements confirm the
prediction~\eqref{SM:eq:Fwilhelmy2}, within a few percent.

As predicted, the surface-tension contribution, $2\gamma_{\rm LG} \cos\phi$, pushes
the plate towards the gas phase. As we show in Section~\ref{SM:sec:LaplacePressure}, $\Delta p$ has contributions not only from an
equilibrium-like Laplace term but also from steady-state
currents. Figure~\ref{SM:fig:Wilhelmy} shows that the overall (small)  contribution of $\Delta p$ is
also to expel the plate from the liquid. Finally, for short
persistence lengths, the currents are negligible
(Fig.~\ref{SM:fig:Wilhelmy}c) and barely contribute to the force on the
plate. On the contrary, at large persistence lengths, large-scale
vortices develop in the vicinity of the plate
(Fig.~\ref{SM:fig:Wilhelmy}d). Interestingly, they oppose the interface
contribution and tend to push the plate back into the liquid. All in
all, the expulsion of the plate by the fluid is an unambiguous
demonstration that the surface tension of active fluids, understood in
its traditional mechanical sense, is indeed negative.

\subsection{Integration of $\sigma_{xy}$ across a liquid-gas interface.}

\begin{figure}[t]
    \centering
    \begin{tikzpicture}[line width=0.3mm, >=latex]
        \draw[dashed] (-4,0) -- (4,0);
        \filldraw (-4,0) circle (0.05) node[left] {$(x_1,y)$};
        \filldraw (4,0) circle (0.05) node[right] {$(x_2,y)$};
        
        \draw (-1.5,-2) -- (1.5,2);
        \draw (0.3,0) arc (0:53.13:0.3) node[right,xshift=0.1cm,yshift=0.1cm] {$\phi(y)$};
        \filldraw (0,0) circle (0.05);

        \node at (-4,2) {Phase 1};
        \node at (4,2) {Phase 2};

        \draw[->] (-0.5+0.25,0+0.33) -- (0.25+0.25,1+0.33) node[above] {$\hat{x}'$};
        \draw[->] (-0.5+0.25,0+0.33) -- (-1.5+0.25,0.75+0.33) node[above] {$\hat{y}'$};

        \draw[->] (3,-2) -- (4.25,-2) node[right] {$\hat{x}$};
        \draw[->] (3,-2) -- (3,-2+1.25) node[right] {$\hat{y}$};

    \end{tikzpicture}
    \caption{\label{SM:fig:int_sigmaxy} To derive Eq.~\eqref{SM:eq:gamma_cos_phi}, we consider an integration path from $(x_1,y)$ to $(x_2,y)$ across an interface parametrized by $(x(s),y(s))$. The interface width $w$ is assumed to be of the order of $\ell_p$, which is much smaller than the radius of curvature of the interface. }
\end{figure}

To derive Eq.~\eqref{SM:eq:Fwilhelmy2} and the active Young-Dupr\'{e} equation, we use the following relation, that we derive below,
\begin{equation}
    \int_{x_1}^{x_2} \sigma_{xy}(x,y)\,dx=\gamma_{\rm LG}\cos\phi(y)\;, \label{SM:eq:gamma_cos_phi}
\end{equation}
where the integration path crosses a liquid-gas interface located at $(x_I,y)$, $x_{1,2}$ are positions in the isotropic bulk of the two phases, and $\phi(y)$ is the angle between the interface and the horizontal axis at height $y$, as shown in Fig.~\ref{SM:fig:int_sigmaxy}. 

\begin{figure}
    \centering
      \begin{tikzpicture}[line width=0.3mm,>=latex]
        \def\x{6}
        \def\y{4}
        \def\dx{-2}
        \def\dy{1}
        \path (-\x,\y) node {\includegraphics{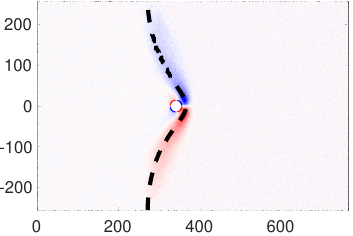}};
        \draw (-\x+\dx,\y+\dy) node[anchor=south west] {\bf a)};
        
        \path (0,\y) node {\includegraphics{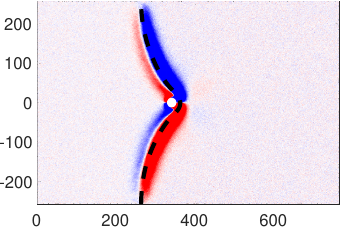}};
        \draw (\dx,\y+\dy) node[anchor=south west] {\bf b)};
        
        \path (\x-2.3,\y+0.13) node {\includegraphics{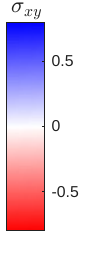}};
      \end{tikzpicture}  
    \caption{$\sigma_{xy}$ measured in the Wilhelmy plate simulations with \textbf{(a)} $v_0=5$, $\tau_p=20/3$, and $\ell_p\simeq 33$, and \textbf{(b)} $v_0=5$, $\tau_p=20$, and $\ell_p=100$. The black dashed curves denote the position of liquid-gas interface defined by $\rho=(\rho_{\rm L}+\rho_{\rm G})/2$. The two panels correspond to the simulations shown in Fig.~\ref{SM:fig:Wilhelmy}a-b, respectively.}
    \label{SM:fig:sigmaxy}
\end{figure}

To derive Eq.~\eqref{SM:eq:gamma_cos_phi}, we first note that $\sigma_{xy}\simeq 0$ in the isotropic bulks and is non-vanishing only near the interface. This statement, based on symmetries, is confirmed by numerical simulations in Fig.~\ref{SM:fig:sigmaxy}. To compute the integral entering Eq.~\eqref{SM:eq:gamma_cos_phi}, we rotate the coordinate system $(x,y)\to(x',y')$ by an angle $\phi(y)$, so that the direction $\hat{x}'$ is tangential to the interface. Furthermore, we assume that the curvature of the interface is small enough that $\sigma$ is locally invariant by translation along the interface, on length scales comparable to the interface width $w$; $\sigma$ is thus independent of $x'$. Using the tensor transformation rule, the stress tensor becomes, in the new coordinate system:
\begin{equation}
    \sigma_{xy}=\sigma_{x'y'}\cos^2\phi+(\sigma_{x'x'}-\sigma_{y'y'})\sin\phi\cos\phi-\sigma_{y'x'}\sin^2\phi\;.
\end{equation}
We then note that $\sigma_{x'y'}=\sigma_{y'x'}=0$ because the system is locally invariant by translation and parity symmetry along the interface, and thus free of any shear. The integral can then be calculated as
\begin{align}\label{SM:eq:gammacosphi}
    \int_{x_1}^{x_2} \sigma_{xy}\,dx=\int_{x_1}^{x_2} (\sigma_{x'x'}-\sigma_{y'y'})\sin\phi\cos\phi\,dx 
    =\cos\phi(y)\int_{y_2'}^{y_1'} (\sigma_{x'x'}-\sigma_{y'y'})\,dy'=\gamma_{\rm LG}\cos\phi(y)\;,
\end{align}
where we have used that $dy'=-\sin\phi\, dx$, that $\phi$ is constant on scales comparable to $w$, and the definition of $\gamma_{\rm LG}$.

\section{Derivation of the active Young-Dupr\'e equation and the liquid-gas interface}\label{SM:sec:ayd}

\begin{figure}
    \centering
    \begin{tikzpicture}[>=latex,line width=0.3mm,scale=2]
        \def\hx{4};
        \def\hy{3};
        \def\cy{0.05};
        \def\offx{0.5};
        \def\figbcx{\offx+0.5*\hx};
        \begin{scope}[xshift=\figbcx cm, yshift=0.3cm]
        \filldraw[blue!15] (0,0) arc(165:111.026:2.96471) -- (1.8,0);
         
        \fill[gray!15] (-1.8,0) rectangle (1.8,-0.5);
        \draw[dashed] (-1.5,-0.5) node[below]{$x_{\rm G}$} -- (-1.5,1);
        \draw[dashed] (1.5,-0.5) node[below]{$x_{\rm L}$} -- (1.5,1);
        \draw[dashed] (-1.5,1) node[left] {$y$} -- (1.5,1);
        \draw[fill] (0.483338,1) circle (0.02);
        \draw[dashed] (0.483338,1) -- (0.483338+0.75,1+1.34687*0.75);
        \draw (0.483338+0.2,1) arc(0:53.4075:0.2) node[right,xshift=0.1cm] {$\phi(y)$};
	\draw (-1.8,0) -- (1.8,0) node[right] {$y=0$};
        \end{scope}
 
    \end{tikzpicture}
    
    \caption{Geometry considered to derive Eq.~\eqref{eq:active_young_dupre} of the main text.}
    \label{SM:fig:contact_angle}
\end{figure}

Here we derive the active Young-Dupr\'e equation from the force balance equation~\eqref{eq:ss_flow} of the main text, also given in Eq.~\eqref{SM:eqn_force_balance_app}. Consider an interface as shown in Fig.~\ref{SM:fig:contact_angle}. We denote $x_{\rm G,L}$ abscissa far away from the interface, such that currents are negligible at $x=x_{\rm G,L}$, and $y$ an ordinate away from the solid surface. We integrate the $x$ component of Eq.~\eqref{SM:eqn_force_balance_app} in the dashed box shown in Fig.~\ref{SM:fig:contact_angle} and note that $\partial_x U_w=0$ to get
\begin{align}
    0=&\int_{x_{\rm G}}^{x_{\rm L}}dx'\int_{-\infty}^{y}dy'\,[\partial_x\sigma_{xx}(\rr')+\partial_y\sigma_{xy}(\rr')-\mu^{-1}J_x(\rr')] \nonumber \\
    =&\int_{-\infty}^{y}dy'\,[\sigma_{xx}(x_{\rm L},y')-\sigma_{xx}(x_{\rm G},y')]+\int_{x_{\rm G}}^{x_{\rm L}}dx'\,\sigma_{xy}(x',y)-\int_{x_{\rm G}}^{x_{\rm L}}dx'\int_{-\infty}^{y}dy'\, \frac{J_{x}(\rr')}{\mu}\;.\label{SM:eq:int_force_balance}
\end{align}
The first integral in Eq.~\eqref{SM:eq:int_force_balance} can be related to fluid-solid surface tensions by subtracting and adding the (constant) pressures $p_{\rm G,L}=-\sigma_{yy}(x_{\rm G/L},y_b)$, where $y_b$ is a position deep in the bulk such that $\sigma_{\alpha\beta}$ is isotropic and homogeneous. This turns the integral into
\begin{align}
    &\int_{-\infty}^{y}dy'[\sigma_{xx}(x_{\rm L},y')-\sigma_{xx}(x_{\rm G},y')] \nonumber \\
    =&\int_{-\infty}^{y}dy'\sigma_{xx}(x_{\rm L},y')+yp_{\rm L}-\int_{-\infty}^{y}dy'\sigma_{xx}(x_{\rm G},y')-yp_{\rm G}+y(p_{\rm G}-p_{\rm L}) \nonumber \\
    =&\gLS-\gGS-y\Delta p\;, \label{SM:eq:int_sigmaxx}
\end{align}
with $\Delta p=p_{\rm L}-p_{\rm G}$ and where we have used the definitions of the fluid-solid surface tensions~\cite{zakine2020}
\begin{equation}
    \gamma_{\rm GS/LS}=y_bp_{\rm G/L}+\int_{-\infty}^{y_b}dy\,\sigma_{xx}(x_{\rm G/L},y)\;.
\end{equation}
We note that these definitions do not depend on $y_b$. Defining the ``drag below $y$''
\begin{equation}
    F_D(y)=-\frac{1}{\mu}\int_{x_{\rm G}}^{x_{\rm L}}dx'\int_{-\infty}^{y}dy'\, J_{x}(\rr') \label{SM:eq:FD}
\end{equation}
and substituting Eqs.~\eqref{SM:eq:gamma_cos_phi},~\eqref{SM:eq:int_sigmaxx}, and~\eqref{SM:eq:FD} into Eq.~\eqref{SM:eq:int_force_balance}, we find
\begin{equation}
    \gLS-\gGS-y\Delta p+\gLG\cos\phi(y)+F_D(y)=0\;, \label{SM:eq:aydex}
\end{equation}
which is Eq.~\eqref{eq:active_young_dupre} of the main text.

\section{Contributions to $\Delta p=p_{\rm L}-p_{\rm G}$ in the presence of a curved interface: the generalized Laplace pressure}
\label{SM:sec:LaplacePressure}
In this section, we compute the contributions entering $\Delta p$ in the presence of a curved interface. We show that, in addition to a term proportional to $\kappa \gamma_{\rm LG}$, where $\kappa$ is the curvature of the interface, there is also a contribution arising from the non-vanishing steady currents.

\begin{figure}[t]
    \centering
    \begin{tikzpicture}[line width=0.3mm, >=latex]
        \draw[dashed] (-4,0) -- (4,0);
        \filldraw (-4,0) circle (0.05) node[left] {$(x_1,y)$};
        \filldraw (4,0) circle (0.05) node[right] {$(x_2,y)$};
        
        \draw[samples=20,domain=1.5:-2] plot(\x,{((\x+3)^2-9)/4.5}) node[left] {$x(s),y(s)$};
        \draw[dashed] (0,0) -- (1.5,2);
        \draw (0.3,0) arc (0:53.13:0.3) node[right,xshift=0.1cm,yshift=0.1cm] {$\phi(y)$};
        \filldraw (0,0) circle (0.05);

        \node at (-4,2) {Phase 1};
        \node at (4,2) {Phase 2};

    \end{tikzpicture}
    \caption{\label{SM:fig:laplacepressure} To derive Eq.~\eqref{SM:eq:generalized Laplace}, we consider an integration path from $(x_1,y)$ to $(x_2,y)$ across an interface parametrized by $(x(s),y(s))$.}
\end{figure}

We start from the force-balance Eq.~\eqref{eq:ss_flow} in the absence of external forces:
\begin{equation}
    \J=\mu\nabla\cdot\bm{\sigma}\;. \label{SM:eq:force_balance_noVw}
\end{equation}
If there are non-vanishing currents in the steady state, Eq.~\eqref{SM:eq:force_balance_noVw} implies that the stress $\bm{\sigma}$ is not uniform, and the same holds for the pressure field $p=-{\rm Tr}\bm{\sigma}/2$. 

Let us calculate the pressure difference $\Delta p$ between two points $(x_{1},y)$ and $(x_{2},y)$ located in the bulk of the liquid and gas phases, respectively {(See Fig.~\ref{SM:fig:laplacepressure})}. Integrating Eq.~\eqref{SM:eq:force_balance_noVw} along the line connecting the two points, we find
\begin{align}
    p_1-p_2&=\frac{1}{\mu}\int_{x_1}^{x_2} J_x(x,y)\,dx-\int_{x_1}^{x_2}\partial_y\sigma_{xy}(x,y)\,dx\\
    &=\frac{1}{\mu}\int_{x_1}^{x_2} J_x(x,y)\,dx-\partial_y \int_{x_1}^{x_2}\sigma_{xy}(x,y)\,dx\;.
\end{align}
Using Eq.~\eqref{SM:eq:gamma_cos_phi}, we get that 
\begin{equation}
    \partial_y \int_{x_1}^{x_2}\sigma_{xy}(x,y)\,dx=\partial_y [\gamma_{\rm LG} \cos\phi(y)]\;,\label{SM:eq:intermediate}
\end{equation}
where we use that the radius of curvature of the interface is much larger than its width so that Eq.~\eqref{SM:eq:gammacosphi} holds. Let us then parameterize the interface as $(x(s),y(s))$, with $s$ being the (signed) arc length, so that
\begin{equation}
-\partial_y\cos\phi(y)=\sin\phi(y)\diff{\phi(y)}{y}=\sin\phi(y)\diff{\phi}{s} \diff{s}{y}=\frac{d\phi}{ds}\;,\label{SM:eq:curvature}
\end{equation}
where we have used that $\sin\phi(y)=y'(s)$ and $ds/dy=1/y'(s)$. Finally, by definition, $d\phi/ds=\kappa$ is the (signed) curvature of the interface. All in all, we find
\begin{equation}
    \int_{x_1}^{x_2}\partial_y\sigma_{xy}(x,y)\,dx=-\kappa\gamma_{\rm LG}\;.
\end{equation}

The term $-\kappa\gamma_{\rm LG}$ is the usual contribution of the Laplace-pressure to the pressure difference between two phases in passive systems. In the presence of steady currents, our computation shows that, in addition, the drag per unit length
\begin{equation}
    \tilde{F}_D\equiv-\frac{1}{\mu}\int_{x_1}^{x_2} J_x(x,y)\,dx
\end{equation}
also enters the force balance through:
\begin{align}\label{SM:eq:generalized Laplace}
    \Delta p&=-\tilde{F}_D+\kappa \gamma_{\rm LG}\;.
\end{align}
Finally, note that in Eq.~\eqref{SM:eq:gamma_cos_phi}, we neglect $\mathcal{O}(\kappa)$ corrections so that there are $\mathcal{O}(\kappa)$ corrections to $\gamma_{\rm LG}\cos(\phi)$ in Eq.~\eqref{SM:eq:intermediate}. Upon differentiating with respect to $y$, these lead to $\mathcal{O}(\kappa^2)$ corrections to Eq.~\eqref{SM:eq:generalized Laplace}. We note that Eq.~\eqref{SM:eq:generalized Laplace} can also be obtained from taking the derivative of Eq.~\eqref{SM:eq:aydex} with respect to a height $y$ in bulk and applying Eq.~\eqref{SM:eq:curvature}.

\begin{figure}
    \centering
  \begin{tikzpicture}
    \def\y{4.3}
    \def\r{.35}    
    \begin{scope}[xshift=-6cm]
    \def\xi{-.25}
    \def\yi{.5}
      \path(0,\y) node{\includegraphics{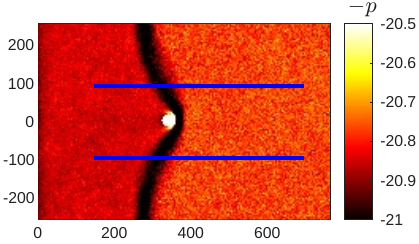}};      

      \path(0,0) node{\includegraphics{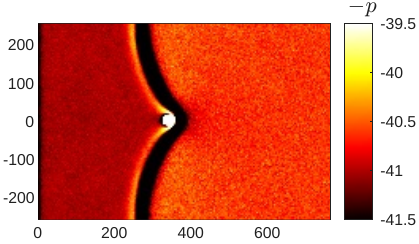}};
      \def\xa{-2.5}
      \def\ya{1.3}
      \def\dw{.3}
      \filldraw[white] (\xa-\dw,\y+\ya-\dw) rectangle (\xa+\dw,\y+\ya+\dw);
      \draw (\xa,\y+\ya) node {\bf (a)};

      \def\xc{-2.5}
      \def\yc{1.3}
      \def\dw{.3}
      \filldraw[white] (\xc-\dw,\yc-\dw) rectangle (\xc+\dw,\yc+\dw);
      \draw (\xc,\yc) node {\bf (b)};
    \end{scope}
    
    \path (1.5,1.5) node{\includegraphics{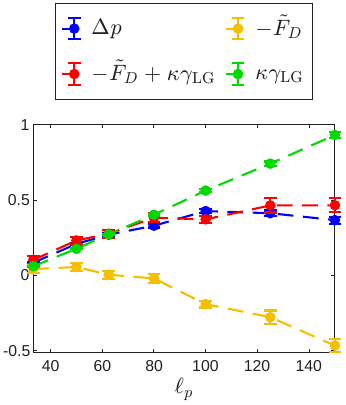}};
    \draw (-0.25,5.45) node {\bf (c)};
  \end{tikzpicture}  
  
    \caption{Contributions to the Laplace pressure $\Delta p=p_{\rm L}-p_{\rm G}$: numerical verification of Eq.~\eqref{SM:eq:generalized Laplace}.  \textbf{(a-b)} Trace of stress tensor fields ${\rm Tr} \bm{\sigma}/2=-p$ measured in the Wilhelmy plate simulations with (a) $v_0=5$ and $\tau_p=20/3$ so that $\ell_p\simeq 33$, and (b) $v_0=5$ and $\tau_p=20$ so that $\ell_p=100$. \textbf{(c)} Measurements of the various contributions entering Eq.~\eqref{SM:eq:generalized Laplace}. The measurements of the bulk pressures $p_{\rm L,G}$ and the surface tension $\gamma_{\rm LG}$ are described in the \hyperref[sec:methods]{Methods}. The drag $\tilde{F}_D$ is integrated along the blue lines, and then averaged over $y\in[-120,-70]\cup [70,120]$. The curvature $\kappa$ is measured by fitting the liquid-gas interfaces with a circle. For small persistence length, the contribution of the drag $\tilde{F}_D$ is negligible whereas it plays an important role for the largest value of $\ell_p$. The simulation parameters correspond to Fig.~\ref{SM:fig:Wilhelmy}. Error bars represent three times the standard error of the mean (s.e.m.), estimated from the same samples as in Fig.~\ref{SM:fig:Wilhelmy}.}
    \label{SM:fig:trsigma}
\end{figure}

In Fig.~\ref{SM:fig:trsigma}, we use the Wilhelmy setup to probe the predictions of Eq.~\eqref{SM:eq:generalized Laplace} quantitatively.

\section{Contact angles for different persistence times and wall stiffnesses}
\begin{figure}[h!]
    \centering
  \begin{tikzpicture}[line width=0.3mm,>=latex]
    \def\x{6}
    \def\y{-4.2}
    \def\dx{-2}
    \def\dy{1.7}
    \path (-\x,0) node {\includegraphics{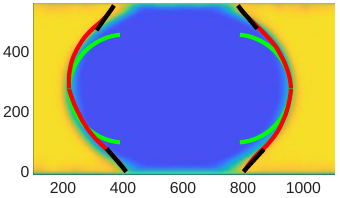}};
    \draw (-\x+\dx,\dy) node[anchor=south west] {\bf a)};
    
    \path (0,0) node {\includegraphics{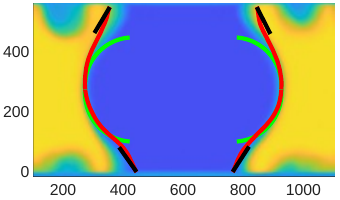}};
    \draw (\dx,\dy) node[anchor=south west] {\bf b)};
    \draw (-1.62*\x,0.15) node {\includegraphics{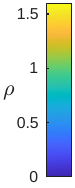}};
    
    \path (-\x,\y) node {\includegraphics{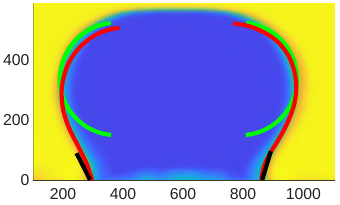}};
    \draw (-\x+\dx,\y+\dy) node[anchor=south west] {\bf c)};
    
    \path (0,\y) node {\includegraphics{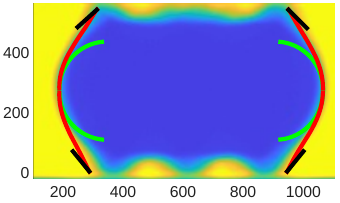}};
    \draw (\dx,\y+\dy) node[anchor=south west] {\bf d)};

    \path (-0.5*\x,1.65*\y) node{\includegraphics{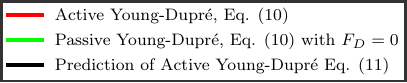}};
    
  \end{tikzpicture}  
    
    \caption{Predictions of interfaces and contact angles using the Active Young-Dupr\'e equation, for different persistence lengths and wall stiffnesses. \textbf{(a)} $\tau_p=12.5$, $\epsilon_w=0.2$. \textbf{(b)} $\tau_p=12.5$, $\epsilon_w=0.1$.  \textbf{(c)} $\tau_p=20$, $\epsilon_w=0.1$ for the bottom wall and $\epsilon_w=2$ for the top wall. \textbf{(d)} $\tau_p=50$, $\epsilon_w=0.1$. The red curves show the predictions of the interfaces, Eq.~\eqref{eq:interface_shape}, obtained from the Active Young-Dupr\'e equation~\eqref{eq:active_young_dupre}, while the green curves correspond to their current-free counterparts. The black lines show the predicted contact angle using Eq.~\eqref{eq:yongfengdupre}.}
    \label{SM:fig:active_YD_sup}
\end{figure}

In Fig.~\ref{SM:fig:active_YD_sup}, we show the predictions of the interface shapes and contact angles for a variety of persistence times and wall stiffnesses, which illustrates the applicability of the active Young-Dupré equation beyond the case detailed in Fig.~\ref{newfig3}.

\section{Contact angles for different particle stiffnesses}

\begin{figure}
    \centering
  \begin{tikzpicture}[line width=0.3mm,>=latex]
    \def\x{4.08}
    \def\y{-4}
    \def\dx{-1.3}
    \def\dy{1.23}
    \path (0,0) node {\includegraphics{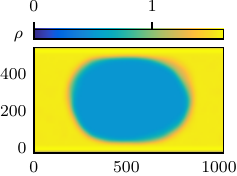}};
    \draw (\dx,\dy) node[anchor=south west] {{\bf a)}};
    
    \path (\x,0) node {\includegraphics{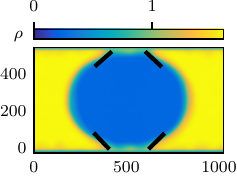}};
    \draw (\x+\dx,\dy) node[anchor=south west] {{\bf b)}};
    
    \path (2*\x,0) node {\includegraphics{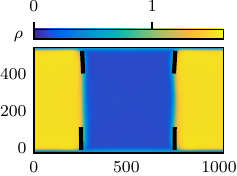}};
    \draw (2*\x+\dx,\dy) node[anchor=south west] {{\bf c)}};
    
    \path (3*\x,0) node {\includegraphics{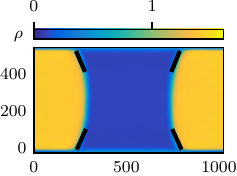}};
    \draw (3*\x+\dx,\dy) node[anchor=south west] {{\bf d)}};
    
    \path (0,\y+0.6) node {\includegraphics{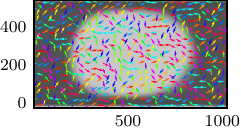}};
    \draw (\dx,\y+\dy+0.5) node[anchor=south west] {\bf e)};
    
    \path (\x,\y+0.6) node {\includegraphics{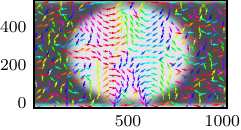}};
    \draw (\x+\dx,\y+\dy+0.5) node[anchor=south west] {\bf f)};
    
    \path (2*\x,\y+0.6) node {\includegraphics{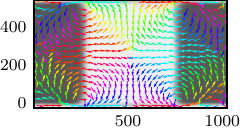}};
    \draw (2*\x+\dx,\y+\dy+0.5) node[anchor=south west] {\bf g)};
    
    \path (3*\x,\y+0.6) node {\includegraphics{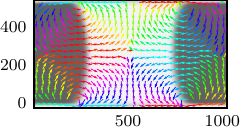}};
    \draw (3*\x+\dx,\y+\dy+0.5) node[anchor=south west] {\bf h)};

    \draw (1.02*\x,1.3*\y) node {\includegraphics{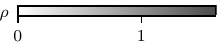}};
    \draw (2.05*\x,1.3*\y) node {\includegraphics{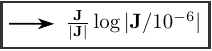}};
    \node at (2.85*\x,1.3*\y) {$\arg\J$};
    \draw (3.2*\x,1.3*\y) node {\includegraphics{colorwheel}};
    
    \def\xx{5.6}
    \def\xoff{0.3}
    
    \path (\xoff,1.9*\y-0.75) node {\includegraphics{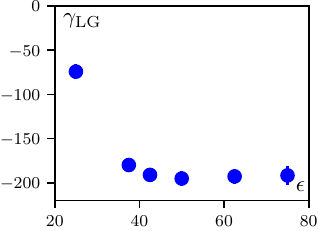}};
    \draw (\dx,1.9*\y+\dy) node[anchor=south west] {{\bf i)}};
        
    \path (\xx+\xoff,1.9*\y-0.75) node {\includegraphics{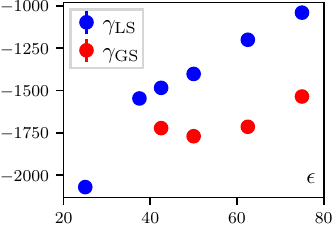}};
    \draw (\xx+\dx+0.1,1.9*\y+\dy) node[anchor=south west] {{\bf j)}};
    
    \path (2*\xx+\xoff,1.9*\y-0.75) node {\includegraphics{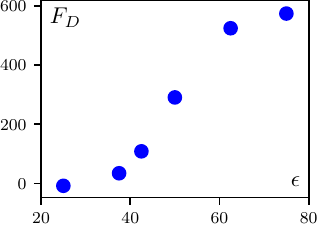}};
    \draw (2*\xx+\dx-0.1,1.9*\y+\dy) node[anchor=south west] {{\bf k)}};
  \end{tikzpicture}  
    
    \caption{Predictions of interfaces and contact angles using the Active Young-Dupr\'e equation for different particle stiffnesses. \textbf{(a,e)} $\epsilon=25$. \textbf{(b,f)} $\epsilon=42.5$. \textbf{(c,g)} $\epsilon=62.5$. \textbf{(d,h)} $\epsilon=75$. Panels (a-d) show the density map of the system, with the black lines showing the contact angles predicted by Eq.~\eqref{eq:yongfengdupre}. Panels (e-h) show the  particle currents, with arrow lengths proportional to $\log|\J/10^{-6}|$. Colors encode the current direction. Panels \textbf{(i-k)} show how the surface tensions and the drag vary as a function of $\epsilon$. $F_D$ is measured separately and averaged in 
    $[0,.5L_x]\times [-\infty,0]$ and $[0,.5L_x]\times [L_y,\infty]$. It is then averaged with $-F_D$ measured in $[.5L_x,L_x]\times [-\infty,0]$ and  $[.5L_x,L_x]\times [L_y,\infty]$. Note that, for $\epsilon=25$, the system is fully wetting and $F_D=0$. Error bars represent three times the standard error of the mean (s.e.m.), estimated from $M\geq 786$ independent block samples (see \hyperref[sec:methods]{Methods}). Parameters: $v=5$, $\tau_p=20$, $\epsilon_w=0.1$, $L_x=1024$, $L_y=512$.}
    \label{SM:fig:active_YD_interaction_stiffness}
\end{figure}

In Fig.~\ref{SM:fig:active_YD_interaction_stiffness}, we vary the stiffness of the particles $\epsilon$ from 25 to 75 with fixed persistence length $\ell_p=100$ and wall stiffness $\epsilon_w=0.1$. Our simulations show that, as in equilibrium, the strength of the interaction between particles affects all surface tensions. Here, the drag is also impacted.

As the particle stiffness is decreased, these variations are such that a  wetting transition is induced: While the liquid phase made by stiff particles only partially wets the solid confining surface, the softest particles fully wet the wall. Consistently, the wetting angle increases with the stiffness of the particles. For partial wetting, the active Young-Dupr\'e equation~\eqref{eq:yongfengdupre} successfully predicts the contact angle.

\section{Stability of bulk phase separation}\label{SM:sec:stability}
In Fig.~\ref{newfig4}, we have shown that an adsorbed droplet of initial radius $R_0=175$ is unstable and spontaneously splits. SM Movie 2 shows an instability to develop in a time of order $\mathcal{O}(10^3\tau_p)$ and that the droplet splits in a time $\sim \mathcal{O}(10^4\tau_p)$. 

In SM Movie 3, we show that bulk droplets of the same initial radius ($R_0=175$) are stable over a time of $3\times 10^5\tau_p$, i.e. 10 times the duration of SM Movie 2. In SM Movie 4, we then show that bulk droplets that are ten times larger ($R_0=1750$) are stable over times comparable to the duration of SM Movie 2 ($3\times 10^4\tau_p$). We do not observe any sign of any form of bulk instability, which shows that the instability of adsorbed droplets has nothing to do with the bulk instabilities leading to microphase separation reported in the literature for other systems~\cite{tjhung2018cluster,shi2020selforganized,caballero2022activity}.

We note that bulk phase-separation is also stable for parameters corresponding to Fig.~\ref{SM:fig:reverse-ostwald}, where adsorbed droplets are shown to experience reverse Ostwald ripening. We show in SM Movie 6 bulk simulations with the same parameters as in Fig.~\ref{SM:fig:reverse-ostwald}d. (These simulations are used to measure the bulk pressure and the liquid-gas surface tension.) These simulations start with a flat interface between liquid and gas regions. The sizes of the liquid phases are much larger than the liquid droplets in Fig.~\ref{SM:fig:reverse-ostwald}. The simulations then run over $\geq 1.5\times 10^4\tau_p$, longer than the duration of SM Movie 5. The phase separated configuration is perfectly stable. 

These movies demonstrate that the size-selection mechanism reported here for adsorbed droplets cannot be directly inferred from established bulk instabilities.

\begin{figure}[p]
    \centering
    \centering
      \begin{tikzpicture}[line width=0.3mm,>=latex]
        \def\x{4}
        \def\y{-3.7}
        \def\dx{-1.5}
        \def\dy{1.5}
        \path (-\x,0) node {\includegraphics{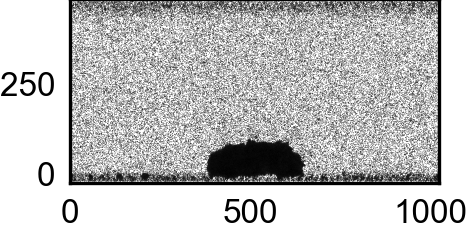}};
        \draw (-\x+\dx,\dy-0.5) node[anchor=south west] {{\bf a)}\ \ \ $t=50\tau_p$};
        
        \path (0,0) node {\includegraphics{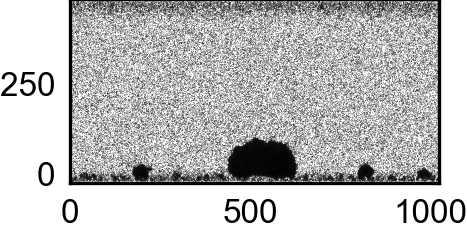}};
        \draw (\dx,\dy-0.5) node[anchor=south west] {{\bf b)}\ \ \ $t=300\tau_p$};
        
        \path (\x,0) node {\includegraphics{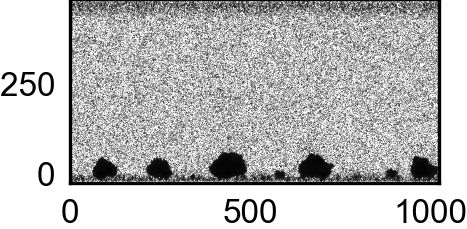}};
        \draw (\x+\dx,\dy-0.5) node[anchor=south west] {{\bf c)}\ \ \ $t=5000\tau_p$};
        
        \path (2*\x+0.25,0) node {\includegraphics{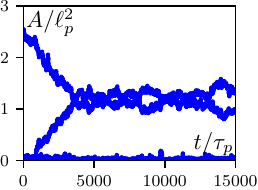}};
        \draw (2*\x+\dx,\dy) node[anchor=south west] {{\bf d)}};
        \node[draw,cross out,ultra thick, rotate=45] at (2*\x+1.7,0.8*\dy) {};
        
        \path (-\x,\y) node {\includegraphics{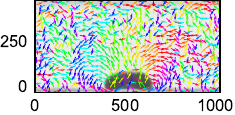}};
        \draw (-\x+\dx,\y+\dy+0.5) node[anchor=south west] {\bf e)};
        \draw (-\x+\dx-0.3,\y+\dy-0.5) node[anchor=south west] {\includegraphics{fig3d_legend}};
        
        \path (0,\y) node {\includegraphics{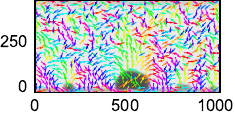}};
        \draw (\dx,\y+\dy+0.5) node[anchor=south west] {\bf f)};
        \draw (\dx+0.2,\y+\dy-0.2) node[anchor=south west] {$\arg\J$};
        \draw (\dx+1.2,\y+\dy-0.5) node[anchor=south west] {\includegraphics{colorwheel}};
        
        \path (\x,\y) node {\includegraphics{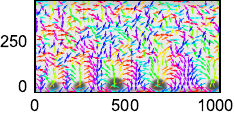}};
        \draw (\x+\dx,\y+\dy+0.5) node[anchor=south west] {\bf g)};
        \draw (\x+\dx-0.34,\y+\dy-0.5) node[anchor=south west] {\includegraphics{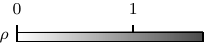}};
        \node[draw,cross out,ultra thick] at (\x+\dx+0.4,\y+\dy-0.8 ) {};
        
        \path (2*\x+0.25,\y) node {\includegraphics{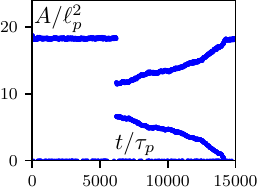}};
        \draw (2*\x+\dx,\y+\dy) node[anchor=south west] {\bf h)};
        \node[draw,regular polygon,regular polygon sides=5] at (2*\x+1.7,\y+0.8*\dy+0.1) {};

        \def\xs{0.15}
        \path (-0.8*\x-\xs-0.05,2.1*\y) node {\includegraphics{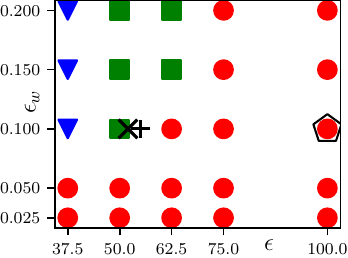}};
        \draw (-0.8*\x-\xs+1.4*\dx-0.05,2.1*\y+1.5*\dy) node[anchor=south west] {\bf i)};

        \path (0.7*\x+\xs,2.1*\y) node {\includegraphics{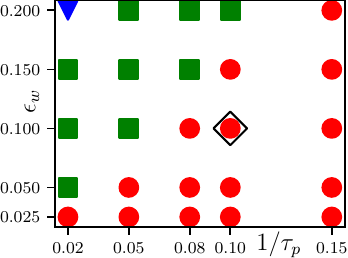}};
        \draw (0.7*\x+\xs+1.4*\dx,2.1*\y+1.5*\dy) node[anchor=south west] {\bf j)};

        \draw (2.06*\x,2*\y) node {\includegraphics{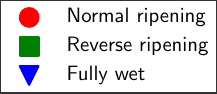}};
      \end{tikzpicture}
\end{figure}
\begin{figure}[p]
    \centering
    \centering
    \caption{Dynamics of adsorbed droplet in the reverse Ostwald ripening regime. \textbf{(a-c)}, Successive snapshots  at (a) $t=50\tau_p$, (b) $300\tau_p$, and (c) $5000\tau_p$. \textbf{(e-g)}, The corresponding current fields. The gray scale shows the particle density field and the arrows represent the current field. Both fields are averaged over $[t-25\tau_p,t+25\tau_p]$ for each time $t$. The arrow lengths are proportional to $\log|\J/10^{-6}|$ and their colors encode the current direction. The interaction strength is $\epsilon=52$. \textbf{(d)}, Time evolution of the surface area $A$ of adsorbed droplets in a system with reverse Ostwald ripening. We adjust the particle stiffness to $\epsilon=55$ so that only one or two droplets are observed throughout the simulation. At $t=0$, the system is initialized with a single large adsorbed droplet of surface area $A_0\simeq \pi R_0^2/2 \simeq 2.45\ell_p^2$. As time goes on, the reversed Ostwald ripening makes a small distant droplet grow from $A\simeq 0$ up to $A\simeq A_0/2$ by absorbing particles that evaporate from the large droplet, eventually leading to two droplets of equal sizes. \textbf{(h)}, Increasing the particle stiffness to $\epsilon=100$ restores the standard Ostwald ripening mechanism. The large droplet is metastable and eventually splits into two large droplets of unequal surface area. Then, the smaller droplet evaporates, leading to the growth of the larger droplet, similar to Fig.~\ref{newfig4}. \textbf{(i-j)} Phase diagrams for droplet dynamics in (i) the $\epsilon$-$\epsilon_w$ plane with fixed $\tau_p=20$ and (j) the $\tau_p$-$\epsilon_w$ plane with fixed $\epsilon=50$. Blue triangles represent fully wetting of the wall by adsorbed droplets. Green squares represent adsorbed droplets undergoing reverse Ostwald ripening. Red circles represent droplets that split following the nucleation of gas bubbles and then experience normal ripening. The parameters corresponding to panels d, g, and h are indicated by the corresponding black symbols. The diamond symbol in panel j corresponds to the active droplet in Fig.~\ref{newfig4} of the main text. Other parameters: $v=5$, $\tau_p=20$, $\epsilon_{w,\rm bottom}=0.1$. In Panels a-g, the droplet initial radius is $R_0=125$ and the system size is $L_x=1024$, $L_y=432$. In Panel h, $R_0=350$, and the system size is $L_x=1536$, $L_y=688$.}
    \label{SM:fig:reverse-ostwald}
\end{figure}

\section{Adsorbed droplets undergoing reverse Ostwald ripening}

In the main text, we show in Fig.~\ref{newfig4} that a large enough active droplet adsorbed on a wall is unstable and undergoes a growing-and-splitting dynamics. The splitting dynamics, shown in SM Movie 2, starts with a bubble nucleation inside the droplet, next to the wall. The bubble then grows in size and the droplet finally splits. Notably, the two resulting droplets then undergo a standard Ostwald ripening: the smaller droplet evaporates, leading to the growth of the larger one. The larger droplet then splits again at a later time.

For a larger persistence length ($\ell_p=100$), a second size-selection mechanism is observed. As shown in Fig.~\ref{SM:fig:reverse-ostwald}a-c and e-g, small droplets nucleate next to the wall, away from the large initial droplet. Then, the larger droplet evaporates while a reversed Ostwald ripening drives the growth of the distant smaller adsorbed droplets, leading to a final state with multiple droplets of similar sizes. 

We note that, for the same simulation parameters, bulk phase separated systems are stable (SM Movie 6) and no sign of reversed Ostwald ripening is observed in the bulk. As for Fig.~\ref{newfig4}, the size-selection mechanism is thus specific to adsorbed droplets and requires the presence of a confining wall.

To best contrast the two size-selection mechanisms, we adjust particle stiffness and persistence length such that, in our large simulation box, we observe either one or two droplets. In Fig.~\ref{SM:fig:reverse-ostwald}d, the reversed Ostwald ripening makes a small distant droplet grow until it absorbs half the particles from the larger initial droplet. In Fig.~\ref{SM:fig:reverse-ostwald}h, the opposite is observed, corresponding to a standard Ostwald ripening mechanism. 

We present two phase diagrams in Fig.~\ref{SM:fig:reverse-ostwald}i,j to show the regions in the parameter space corresponding to the two regimes. The phase diagrams are obtained by manually observing the dynamics of a single droplet of varying size for each point in the parameter space. We construct the diagram using a system size of $L_x=1024$ and $L_y=432$, while the stability of larger droplets is tested using a larger system with $L_x=1536$ and $L_y=688$. We find that single droplets in the reverse Ostwald ripening regime are unstable at any size, and that small distant droplets always nucleate and grow. This is used to distinguish the reverse and normal ripening regimes. The phase diagrams reveal that the transition to reverse ripening is triggered by increasing persistence, increasing wall stiffness, or decreasing particle stiffness. A fully wetting regime appears in the limit of large persistence, stiff walls, and soft particles. The reverse ripening regime appears to be an intermediate regime along the transition from partial to full wetting.
 
\section{Finite-size effects}\label{SM:sec:finitesize}
\begin{figure*}
\centering
  \begin{tikzpicture}[line width=0.3mm,>=latex]
    \def\x{7}
    \def\y{4.8}
    \def\dx{-2}
    \def\dy{2.2}
    \path (-\x,\y) node {\includegraphics{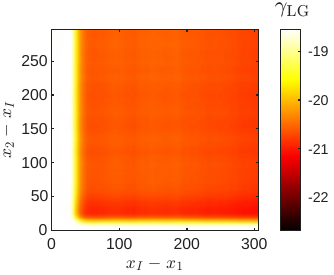}};
    \draw (-\x+\dx,\y+\dy) node[anchor=south west] {\bf a)};
    
    \path (0,\y) node {\includegraphics{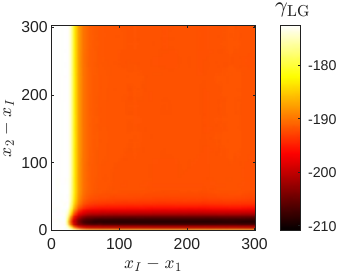}};
    \draw (\dx,\y+\dy) node[anchor=south west] {\bf b)};
    
  \end{tikzpicture}  
\caption{$\gamma_{\rm LG}(x_1,x_2)=\int_{x_1}^{x_2} (\sigma_{yy}(x)-\sigma_{xx}(x))dx$. \textbf{(a)} $\tau_p=20/3$. \textbf{(b)} $\tau_p=20$. The systems are periodic in both directions.}
  \label{SM:fig:heatmap_LG_tension}
\end{figure*}

\begin{figure}
\centering
  \begin{tikzpicture}[line width=0.3mm,>=latex]
    \def\x{7}
    \def\y{3.8}
    \def\dx{-2}
    \def\dy{1.5}
    \path (-\x,\y) node {\includegraphics{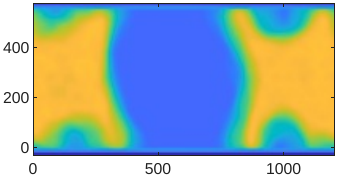}};
    \draw (-\x+\dx,\y+\dy) node[anchor=south west] {\bf a)};
    \draw (-1.5*\x,\y+0.29) node {\includegraphics{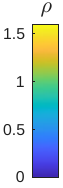}};
    
    \path (0,\y+0.2) node {\includegraphics{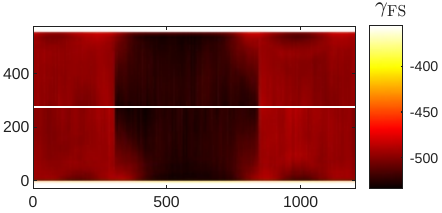}};
    \draw (\dx,\y+\dy) node[anchor=south west] {\bf b)};
    
    \def\dy2{1.5}
    \path (-\x,0) node {\includegraphics[scale=1]{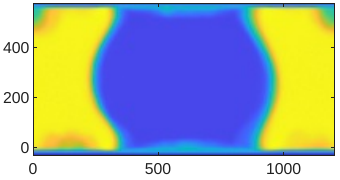}};
    \draw (-\x+\dx,\dy2) node[anchor=south west] {\bf c)};
    \draw (-1.5*\x,0.29) node {\includegraphics{colorbar_rho_vertical}};
    
    \path (0,0.2) node {\includegraphics[scale=1]{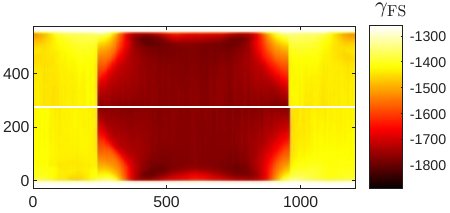}};
    \draw (\dx,\dy2) node[anchor=south west] {\bf d)};
    
  \end{tikzpicture}  
\caption{\textbf{(a) and (c)} The density field with $\tau_p=20/3$ (a) and $\tau_p=20$ (c) showing formation of many bubbles near the walls. \textbf{(b) and (d)} The heat map of measuring fluid-solid surface tension $\gamma_{\rm FS}(x,y)=\int_{-\infty}^y \sigma_{xx}(x,y')dy'+yp_{\rm F}$, where $p_{\rm F}=p_{\rm L/G}$ if $x$ falls into the liquid/gas bulk, corresponding to panels (a) and (c).  }
  \label{SM:fig:heatmap_tension}
\end{figure}

\begin{figure}
    \centering
    \includegraphics{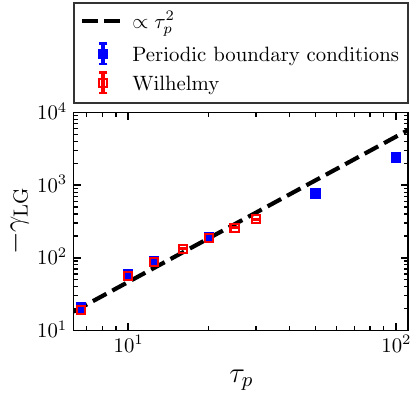}
    \caption{Measurements of $\gamma_{\rm LG}$ in a phase-separated periodic system and in the Wilhelmy plate setups. The finite-size corrections to $\gamma_{\rm LG}$ are of the order of the interface curvature, which scales as $L_y^{-1}$ and are thus completely negligible. Error bars represent three times the standard error of the mean (s.e.m.), estimated from $M\geq 2518$ independent block samples for the periodic systems and $M\geq 8000$ for the Wilhelmy plates (see \hyperref[sec:methods]{Methods}).}
    \label{SM:fig:bulk_vs_wilhelmy}
\end{figure}

To measure bulk observables, we need to make sure that measurements are carried out at distance $\ell \gg \ell_p$ from the interfaces. Figs.~\ref{SM:fig:sigmaxy}a-b for instance show the local stress anisotropy close to interfaces, which is indeed vanishing in bulk phases. Fig.~\ref{SM:fig:trsigma}a-b show how the pressure field becomes homogeneous away from the liquid-gas interface in the Wilhelmy plate experiments. Similarly, Fig.~\ref{SM:fig:heatmap_LG_tension} shows that the measurement of the liquid-gas surface tension in phase-separated systems leads to uniform predictions as long as the integration of the $\sigma_{yy}-\sigma_{xx}$ goes over the full interface region. Next, Fig.~\ref{SM:fig:heatmap_tension}a-d show a standard difficulty met in numerical simulations: long-lived gas and liquid bubbles can nucleate close to the walls in the liquid and gas phases, respectively. These bubbles time-average out but this demands very long simulations. In the figure, we show that, despite the presence of these bubbles, the measurement of the fluid-solid surface tension leads to relatively uniform estimates in the bulk phases, as long as $\gamma_{\rm LS}$ and $\gamma_{\rm GS}$ are measured at distances $\ell\gg\ell_p$ from the walls. Finally, we compare in Fig.~\ref{SM:fig:bulk_vs_wilhelmy} the measurements of the liquid-gas surface tension using Eq.~\eqref{eq:Ftot2} for the Wilhelmy plate setup at $y=\pm L_y/2$ or using phase-separated systems with periodic boundary conditions, leading to negligible differences.

\section{Active Ornstein-Uhlenbeck particles}\label{SM:sec:AOUP}

In this section, we extend our theory to active Ornstein-Uhlenbeck particles (AOUPs) interacting via pairwise forces. We follow~\cite{martin2021statistical}  to derive the hydrodynamic theory and the effective stress tensor. The dynamics of AOUPs is given by the set of Langevin equations:
\begin{eqnarray}
    \dot\rr_i&=&\vv_i-\mu\sum_{j\neq i}\nabla_{\rr_i}U(\rr_i-\rr_j)-\mu\nabla_{\rr_i}U_w(\rr_i)\;, \\
    \tau_p\dot\vv_i&=&-\vv_i+v_0\sqrt{\tau_p}\bm{\xi}_i\;,
\end{eqnarray}
where $\vv_i$ is the self-propelling velocity of particle $i$, and $\bm{\xi}_i$ is a Gaussian white noise with zero mean and unit variance. The other notations are the same as in the main text. The one-body density function $\Psi(\rr,\vv,t)=\langle\sum_i\delta(\rr-\rr_i(t))\delta(\vv-\vv_i(t))\rangle$ then evolves according to:
\begin{eqnarray}
\pdiff{\Psi}{t}&=&\nabla_{\rr}\cdot\left(-\vv\Psi+\mu\int d\rr' d\vv'\,\Psi^{(2)}(\rr,\vv,\notag\rr',\vv',t)\nabla_{\rr}U(\rr-\rr')+\mu\Psi\nabla_{\rr}U_w(\rr)\right)\\&&+\frac{1}{\tau_p}\nabla_{\vv}\cdot(\vv \Psi)+\frac{v_0^2}{2\tau_p}\Delta_{\vv}\Psi\;, \label{SM:eqn_FK_AOUP_1}
\end{eqnarray}
where we have introduced the two-body density function:
\begin{equation}
    \Psi^{(2)}(\rr,\vv,\rr',\vv',t)=\left\langle\sum_{i\neq j}\delta(\rr-\rr_i(t))\delta(\vv-\vv_i(t))\delta(\rr'-\rr_j(t))\delta(\vv'-\vv_j(t))\right\rangle\;.
\end{equation}
The density, orientation and nematic-order fluctuating fields are defined as
\begin{align}
    &\hat{\rho}(\rr,t)=\sum_i\delta(\rr-\rr_i(t))\;,\ \hat{\mm}(\rr,t)=\sum_i\vv_i\delta(\rr-\rr_i(t))\;, \\
    &\hat{\Q}(\rr,t)=\sum_i\left[\vv_i(t)\otimes\vv_i(t)-\frac{1}{d}\vv_i^2\mathbb{I}\right]\delta(\rr-\rr_i(t))\;,
\end{align}
where $\mathbb{I}$ is the identity matrix. Their statistical averages are denoted by $\rho$, $\mm$, and $\Q$, respectively. 

Integrating Eq.~\eqref{SM:eqn_FK_AOUP_1} with respect to $\vv$, we obtain a continuity equation for $\rho$:
\begin{equation}
    \pdiff{\rho}{t}=-\partial_\alpha J_\alpha\;,
\end{equation}
where the particle current is given by
\begin{align}
J_\alpha=&m_\alpha-\mu\int d\rr'\,\rho^{(2)}(\rr,\rr',t)\partial_\alpha U(\rr-\rr')-\mu\rho\partial_\alpha U_w(\rr) \label{SM:eqn_J_two_points_AOUP} \\
=&m_\alpha+\mu\partial_\beta\sigma^{\rm IK}_{\alpha\beta}-\mu\rho\partial_\alpha U_w(\rr)\;,  \label{SM:eqn_J_AOUP}
\end{align}
with $\rho^{(2)}(\rr,\rr',t)$  the two-particle density and $\sigma^{\rm IK}$ the Irving-Kirkwood stress tensor, as defined in Eqs.~\eqref{SM:eqn_density_two_points} and~\eqref{SM:eqn_IK_stress}, respectively. To write $m_\alpha$ as the divergence of an active stress tensor, we calculate the dynamics of $m_\alpha$ by applying the operator $\int \vv d\vv$ on both sides of Eq.~\eqref{SM:eqn_FK_AOUP_1}. We find
\begin{align}
    \pdiff{m_\alpha}{t}=&\partial_\beta \left(-Q_{\alpha\beta}+\mu\int d\rr'\partial_\beta U(\rr-\rr')\left\langle\sum_{i=1}^N\sum_{k\neq i}v_{i,\alpha}\delta(\rr-\rr_i(t))\delta(\rr'-\rr_k(t))\right\rangle\right) \nonumber \\
    &-\frac{1}{d}\partial_\alpha K+\mu\partial_\beta(m_\alpha\partial_\beta U_w)-\frac{1}{\tau_p}m_\alpha\;,
\end{align}
where
\begin{equation}
    K(\rr,t)\equiv \int \vv^2\Psi(\rr,\vv,t)\,d\vv\;.
\end{equation}
Similar to Eq.~\eqref{SM:eqn_m_dynamic}, we can then rewrite the dynamics of $\mm$ as
\begin{align}
\pdiff{m_\alpha}{t}=&\partial_\beta\left(-Q_{\alpha\beta}-\mu I^{(1)}_{\alpha\beta}-\mu I^{\rm ext}_{\alpha\beta}-\frac{K}{d}\delta_{\alpha\beta}\right)-\frac{m_\alpha}{\tau_p}\;,\label{SM:eqn_m_dynamic_AOUP}
\end{align}
where we define for clarity:
\begin{equation}
I^{(1)}_{\alpha\beta}=-\int d\rr'\,\partial_{\beta}U(\rr-\rr')\left\langle \sum_{i=1}^N\sum_{k\neq i}v_{i,\alpha}\delta(\rr-\rr_i(t))\delta(\rr'-\rr_k(t))\right\rangle\;.
\end{equation}

\begin{figure}[h]
    \centering
    \begin{tikzpicture}[line width=0.3mm,>=latex]
    \def\x{8}
    \def\y{4.8}
    \def\dx{-3.2}
    \def\dy{3}
    \path (-\x,\y) node {\includegraphics{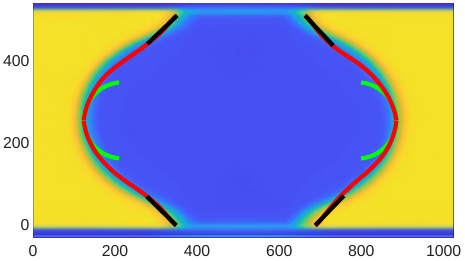}};
    \draw (-\x+\dx,\y+\dy) node[anchor=south west] {\bf a)};
    
    \path (0,\y) node {\includegraphics{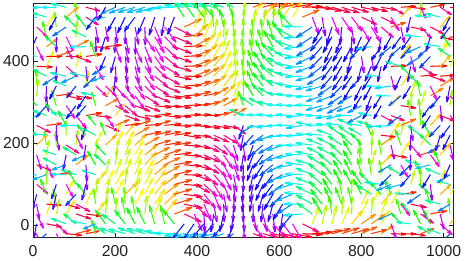}};
    \draw (\dx,\y+\dy) node[anchor=south west] {\bf b)};
    \draw (-0.97*\x,0.3*\y) node {\includegraphics{figS7_legend}};
    \draw (-\x,\y+\dy-0.2) node {\includegraphics{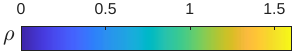}};
    \draw (0,\y+\dy-0.2) node {\includegraphics{fig3d_legend}};
    \draw (-0.8,0.3*\y) node {$\arg\J$};
    \draw (0.8,0.3*\y) node {\includegraphics{colorwheel}};
    \end{tikzpicture}
    \caption{Partial wetting of a confining wall by active Ornstein-Uhlenbeck particles. We present \textbf{(a)} the density field and  \textbf{(b)} the current field of a bulk phase-separated system in the presence of an external confinement along $\hat y$.  The red curves in panel (a) represent the shape of liquid-gas interfaces predicted by the active Young-Dupr\'e equation, while the green curve shows the current-free prediction, obtained by setting $F_D=0$. Black lines show the predicted contact angle using Eq.~\eqref{eq:yongfengdupre}.  We measure $\gamma_{\rm LG}=-157$ in this system, which is negative and consistent with $\gamma_{\rm LG}=-175$ measured in a periodic system. Parameters: $v_0=5$, $\tau_p=12.5$, $\epsilon=50$, $\epsilon_w=0.1$, $L_x=2L_y=1024$.}
    \label{SM:fig:AOUP}
\end{figure}

In the steady state, we have $\partial_t\rho=\partial_t m_\alpha=0$, leading to $\partial_\alpha J_\alpha=0$ and
\begin{equation}
    m_\alpha=\tau_p\partial_\beta\left(-Q_{\alpha\beta}-\mu I^{(1)}_{\alpha\beta}-\mu I^{\rm ext}_{\alpha\beta}-\frac{K}{d}\delta_{\alpha\beta}\right)\;.
\end{equation}
Substituting into Eq.~\eqref{SM:eqn_J_AOUP}, we find that
\begin{equation}
    \frac{1}{\mu}J_\alpha=\partial_\beta\left(-\frac{\tau_p K}{d\mu}\delta_{\alpha\beta}-\frac{\tau_p}{\mu}Q_{\alpha\beta}-\tau_pI^{(1)}_{\alpha\beta}-\tau_pI^{\rm ext}_{\alpha\beta}+\sigma^{\rm IK}_{\alpha\beta}\right)-\rho\partial_\alpha U_w(\rr)\;,
\end{equation}
which shares the same structure as the force balance equation~\eqref{SM:eqn_force_balance_app}. We then identify the  stress tensor as
\begin{equation}
    \sigma_{\alpha\beta}=\sigma_{\alpha\beta}^a+\sigma_{\alpha\beta}^{\rm IK}=-\frac{\tau_p K}{d\mu}\delta_{\alpha\beta}-\frac{\tau_p}{\mu}Q_{\alpha\beta}-\tau_pI^{(1)}_{\alpha\beta}-\tau_pI^{\rm ext}_{\alpha\beta}+\sigma^{\rm IK}_{\alpha\beta}\;,\label{SM:eqn_stress_AOUPs}
\end{equation}
where $\sigma^a_{\alpha\beta}$ denotes the active contributions to the stress tensor. We can then use the definitions of surface tension given in the text in terms of $\sigma$ to extend the Active Young-Dupr\'e equation to AOUPs. Figure~\ref{SM:fig:AOUP} shows our theory to correctly predict  the contact angle and liquid-gas interface of macroscopically phase-separated AOUPs. 

We simulate AOUPs using Heun's method~\cite{zakine2020} and follow a similar protocol as for Figs.~\ref{newfig2} and~\ref{newfig3} of the main text. We measure the tensors in Eq.~\eqref{SM:eqn_stress_AOUPs} as:
\begin{align}
    K(x_i,y_j)&=\frac{1}{r_0^2}\left\langle\sum_{k\sim (i,j)} v_{k}^2\right\rangle\;, \\
    I^{(1)}_{\alpha\beta}(x_i,y_j)+I^{\rm ext}_{\alpha\beta}(x_i,y_j)&=\frac{1}{r_0^2}\left\langle\sum_{k\sim (i,j)} f_{k,\beta} v_{k,\alpha}\right\rangle\;, \\
    Q_{\alpha\beta}(x_i,y_j)&=\frac{1}{r_0^2}\left\langle\sum_{k\sim (i,j)} \left(v_{k,\alpha}v_{k,\beta}-\frac{v_{k}^2}{2}\delta_{\alpha\beta}\right)\right\rangle\;,
\end{align}
where $v_k^2=\vv_k\cdot\vv_k$ and $k$ labels the particles in the cell $(i,j)$. Note that the definition of $\sigma_{\alpha\beta}^{\rm IK}$ is the same as for ABPs.

\end{document}